\documentclass{aa} 
\usepackage{graphicx}
\usepackage{graphicx,amssymb,amsmath,times}
\usepackage[]{natbib}

\usepackage[varg]{txfonts}
\usepackage{booktabs}
\usepackage{url}
\usepackage{xcolor}
\usepackage{sidecap}
\usepackage{amssymb}
\usepackage{soul}

\usepackage{caption}
\usepackage{subcaption}

\newcommand{\Teff}{\ensuremath{T_\mathrm{eff}}}

\newcommand{\logg}{\ensuremath{\log g}}
\def\kms{$\mathrm{km\, s^{-1}}$}

\newcommand{\feh}{[Fe/H]}

\newcommand{\meanks}{$\langle K_S \rangle$}
\newcommand{\meanh}{$\langle H \rangle$}
\newcommand{\meanj}{$\langle J \rangle$}

\bibpunct{(}{)}{;}{a}{}{,} 

\begin{document}

   \title{Using classical Cepheids to study the far side of the Milky Way disk \thanks{Based on observations collected at the European Southern Observatory under ESO programs 095.B-0444(A), 179.B-2002.}}
    \subtitle{I. Spectroscopic classification and the metallicity gradient}

   \author{J.~H.~Minniti\inst{1,2,3}
   \and
   L.~Sbordone\inst{2}
   \and
   A.~Rojas-Arriagada\inst{1,3}
   \and
   M.~Zoccali\inst{1,3}
   \and
   R.~Contreras Ramos\inst{1,3}
   \and
   D.~Minniti\inst{3,4,5}
   \and
   M.~Marconi\inst{6}
   \and
   V.~F.~Braga\inst{3,4}
   \and
   M.~Catelan\inst{1,3}
   \and
   S.~Duffau\inst{4}
   \and
   W.~Gieren\inst{3,7}
   \and
   A.~A.~R.~Valcarce\inst{8,9,10}
   }
   \institute{Pontificia Universidad Cat{\'o}lica de Chile, Instituto de Astrof{\'i}sica, Av. Vicu\~na Mackenna 4860, 7820436, Macul, Santiago, Chile\\
              \email{jhminniti@uc.cl}
         \and
             European Southern Observatory, Alonso de C{\'o}rdova 3107, Santiago, Chile
         \and 
             Millennium Institute of Astrophysics, Av. Vicu\~na Mackenna 4860, 7820436, Macul, Santiago, Chile
         \and
                         Departamento de F{\'i}sica, Facultad de Ciencias Exactas, Universidad Andr{\'e}s Bello, Fern{\'a}ndez Concha 700, Las Condes, Santiago, Chile
          \and
             Vatican Observatory, V00120 Vatican City State, Italy
          \and
             INAF-Osservatorio Astronomico di Capodimonte, Via Moiariello 16, 80131 Naples, Italy
          \and
             Universidad de Concepción, Departamento Astronomía, Casilla 160-C, Concepción, Chile
          \and
             Departamento de F{\'i}sica, Universidade Estadual de Feira de Santana, Av. Transnordestina, S/N, CEP 44036-900 Feira de Santana, BA, Brazil
          \and
             Centro de Nanotecnolog{\'i}a Aplicada, Facultad de Ciencias, Universidad Mayor, Santiago, Chile
          \and
             Centro para el Desarrollo de la Nanociencia y la Nanotecnolog{\'i}a (CEDENNA), Santiago, Chile
             }

   \date{Received ; accepted }

 
  \abstract
   {Much of what we know about the Milky Way disk is based on studies of the solar vicinity. The structure, kinematics, and chemical composition of the far side of the Galactic disk, beyond the bulge, are still to be revealed.}
   {Classical Cepheids (CCs) are young and luminous standard candles. We aim to use a well-characterized sample of these variable stars to study the present-time properties of the far side of the Galactic disk.}
   {A sample of 45 Cepheid variable star candidates were selected from near-infrared time series photometry obtained by the VVV survey. We characterized this sample using high quality near-infrared spectra obtained with VLT/X-Shooter. The spectroscopic data was used to derive radial velocities and iron abundances for all the sample Cepheids. This allowed us to separate the CCs, which are metal rich and with kinematics consistent with the disk rotation, from type II Cepheids (T2Cs), which are more metal poor and with different kinematics.}
   {We estimated individual distances and extinctions using VVV photometry and period-luminosity relations, reporting the characterization of 30 CCs located on the far side of the Galactic disk, plus 8 T2Cs mainly located in the bulge region, of which 10 CCs and 4 T2Cs are new discoveries. The remaining seven stars are probably misclassified foreground ellipsoidal binaries. This is the first sizeable sample of CCs in this distant region of our Galaxy that has been spectroscopically confirmed. We use their positions, kinematics, and metallicities to confirm that the general properties of the far disk are similar to those of the well-studied disk on the solar side of the Galaxy. In addition, we derive for the first time the radial metallicity gradient on the disk's far side. Considering all the CCs with $R_{\mathrm{GC}} < 17\,\rm{kpc}$, we measure a gradient with a slope of $-0.062 \, \mathrm{dex\, kpc^{-1}}$ and an intercept of $+0.59 \, \rm{dex}$, which is in agreement with previous determinations based on CCs on the near side of the disk.}
   {}


  \maketitle

\section{Introduction}

Our own Galaxy, the Milky Way (MW), is the one large spiral whose structure, kinematics, and stellar populations can, at the moment, be studied to the greatest detail. As such, its role in shaping our understanding of spiral galaxies in general is unparalleled. However, our location at $\sim20\,\rm{pc}$ above its disk mid-plane \citep[see][and references therein]{bennett2019} also generates a number of challenges, especially in studying its innermost regions, and those lying beyond them in the disk on the other side of the Galactic center. These difficulties are mostly due to three factors compounding each other: extreme extinction close to the inner disk mid-plane, extreme crowding, and the difficulty obtaining accurate distances for the studied objects. This, in combination with the fact that we are looking through a superposition of structures and stellar populations belonging to different components of the Galaxy, has prevented us from getting a definitive picture of its constitution.

The structures forming the inner part of the MW have proven difficult to disentangle, partly because of the factors mentioned, partly because they are intrinsically complex. However, thanks to a number of dedicated photometric (e.g., VVV and VVVX, \citealt{minniti2010,minniti2016,borissova2018}) and spectroscopic surveys (e.g., APOGEE, \citealt{eisenstein2011}; GIBS, \citealt{zoccali2017}), an increasingly clear picture is developing. The structure called the Bulge (we use this term for simplicity) appears to be a boxy/peanut bulge, including a thick bar, a spheroid, and a long thin bar \citep[for a review, see][]{barbuy2018}. The central thick bar, sometimes called the main bar, is seen from an angle of roughly 27 degrees with respect to the Sun-Galactic center line \citep{wegg2013}. The bulge populations appear to be prevalently old \citep[older than $\sim 7.5-10$\,Gyr, see][]{gennaro2015,surot2019}, but rather metal rich and $\alpha$-rich \citep{barbuy2018}. On the other hand, the most metal-poor population, as identified by RR Lyrae variables, does not seem to follow the boxy/peanut distribution and kinematics, and might be the faint remnant of a true spheroidal bulge or the densest part of the inner halo \citep{minniti1996,dekany2013,dekany2018}. 

The most massive component of the MW, the disk ($4\times 10^{10} M_\sun$, \citealt{blandhawthorn16}, versus $2\times 10^{10} M_\sun$ for the Bulge, \citealt{valenti2016}), is itself recognized as composed by at least two chemically and kinematically different stellar components: the more massive, younger, and more metal-rich thin disk, and the lower metallicity, older, and less massive thick disk. Their relationship is still being debated, and many of their fundamental parameters are still uncertain, including their density scale length and composition gradient \citep{blandhawthorn16}. The interpretation of the disk abundance gradients, moreover, presents challenges of its own since the disk is subject to a number of poorly constrained secular processes, from the dynamical instability that generates the bar to the radial migration of stellar populations (see \citealt{anders2017}, \citealt{toyuchi2018}, \citealt{lemasle2018}, \citealt{maciel2019}). 

In this paper we focus on classical Cepheids (CCs): young ($10-300\,\mathrm{Myr}$, see \citealt{bono2005}) pulsating variable stars with periods ranging mainly between 1 and 100 days, that follow accurate period-luminosity (PL) relations. They can be used to trace the young stellar populations present within the Galactic thin disk and study the structural parameters of this component. In particular, CCs have been used to study the present-day metallicity distribution of the Galactic thin disk (\citealt{genovali2014} and references therein). The radial metallicity gradient provides valuable constraints to models of formation and chemical evolution of the disk. In this work we classify a sample of candidate CCs on the opposite side of the disk with respect to the Galactic center, and use them to determine the metallicity gradient. In a companion paper (Minniti et al., in preparation) we use these CCs to extend our knowledge of the spiral structure towards this largely unknown region of our Galaxy.

The ESO Public Survey VISTA Variables in the Vía Láctea \citep[VVV,][]{minniti2010} is a near-infrared (near-IR) photometric survey that mapped the Galactic bulge and the southern mid-plane in $ZYJHK_S$ from 2010 to 2016 and was then extended \citep[VVVX,][]{vvvxminniti2018} to 2019. In the $K_S$ band it obtained multi-epoch observations, allowing us to search for variable stars in the highly reddened regions of the MW that it mapped.

Recently there have been photometric searches of CCs conducted in the near-IR towards the highly obscured areas of the bulge along the Galactic plane. \citet{matsunaga2011} found three CCs located in the nuclear stellar disk; they reported a fourth one also belonging to this structure of the Galaxy in \citet{matsunaga2015}. Regarding the far side of the disk, \citet{dekany2015a,dekany2015b} reported 37 new candidate CCs at distances greater than about 8\,kpc from the Sun lying close to the Galactic plane that were found using VVV data. These data were interpreted as evidence of the presence of a young thin disk in the central region of the Galaxy. The distances calculated in this work were recalculated by \citet{matsunaga2016} using a different reddening law. They conclude that there is a lack of CCs in the inner part of the Galaxy, and added 18 new objects to the sample of CCs beyond the bulge. 

The studies mentioned above rely on a photometric classification to isolate Cepheids on the far side of the Galaxy. A first selection based only on periods, amplitudes, and light curve shapes yields a sample of Cepheids. However, they come in two varieties that overlap in period and in amplitude. Classical Cepheids are young, massive, core helium burning stars crossing the instability strip while on the blue loops. On the contrary, type II Cepheids (T2Cs) are old, low mass, less luminous population II giants that undergo helium-shell burning. They may cross the instability strip at different stages during their evolution from the blue horizontal branch towards the asymptotic giant branch (AGB) and post-AGB phases \citep[for a review, see][]{catelan2015}. While the absolute luminosities of the two types of Cepheids are very different, their periods and amplitudes largely overlap, and the shapes of the near-IR light curves are not markedly different, which prevents us from making an unambiguous classification in many cases.

Previous studies focusing on CCs towards the inner regions of the Galaxy, relied on the use of a combination of PL relations in two bands in order to determine individual distances and reddenings, assuming that the targets are either CCs or T2Cs \citep[e.g.,][]{matsunaga2011,dekany2015a,dekany2015b}. The reddenings derived under each of the two hypotheses are then compared with reddening maps, and the most plausible is adopted as the true one, together with its associated distance and Cepheid classification. This method is not free from cross-contamination between the two types of variables since reddening can be very patchy and the available maps have limited spatial resolution. At the time of writing, a new classification method for Cepheid near-IR light curves based on a convolutional neural network has been presented by \citet{dekany2019}. Even though this method may be an improvement compared to the one described above, it is also prone to misclassification.

In addition, in these regions highly affected by extinction, the reddening law has been found to be significantly different from the value commonly adopted elsewhere in the Galaxy \citep{nishiyama2006,nishiyama2009,alonso-garcia2017,dekany2019}. Moreover, the total-to-selective extinction ratios determined in these works do not agree. This can lead to differences of up to a few kiloparsec on the derived distances depending on the adopted value, as discussed by \citet{matsunaga2016}.

A clean sample of CC is needed to characterize the chemical and structural properties of the far disk. One possibility to improve this selection, and to minimize the effect of the above-mentioned issues present in the near-IR photometric classification, is to add more information. In particular, obtaining the metallicity and kinematics by means of spectroscopy would allow us to better separate the two classes of variables. Due to their old ages, T2Cs are expected to be found preferentially in the bulge or halo rather than the disk, and therefore their kinematics would differ from that of CCs. They are also expected to have significantly lower metallicities ($\mathrm{\feh}\sim-1$ or lower) than typical thin disk stars ($\mathrm{\feh}\sim0$). At the same time, combining their photometric and spectroscopic information, we can study the metallicity gradient and kinematics on the far side of the MWs disk. Moreover, a well-characterized sample of Cepheids can be used to further validate the mentioned near-IR classification methods.

In a pioneering work, \citet{inno2019} combined near-IR photometry with spectroscopic information for the first time to classify five Cepheids. In this paper we do this for a sample nearly ten times larger using higher resolution near-IR spectra.

This paper is organized as follows. Section~\ref{section:photometry} presents the near-IR photometry of the sample Cepheids, including the ephemerides, extinction law, and distance determinations. Section~\ref{section:spectroscopy} describes the spectroscopic observations and data reduction. Section~\ref{section:kinematics} presents the radial velocity measurements and the kinematics for the sample stars. In Sect.~\ref{section:classification} we discuss the spectroscopic determination of the fundamental atmospheric parameters, using full spectral fitting and synthetic models, the Cepheids classification, and the metallicity gradient on the far side of the MW. In Sect.~\ref{section:summary} we summarize our findings. 

\section{Photometry}\label{section:photometry}

The candidate CCs on the far side of the Galactic plane were selected based on the VVV Survey near-IR photometry of the Galactic plane in the Galactic longitude range $-40$\degr$ < l < +10$\degr. The VVV Survey is a near-IR variability survey mapping the main components of the MW (i.e., the bulge and the southern part of the Galactic disk). The multi-epoch observations are carried out in the $K_S$ band, with a total of $\sim$70 epochs obtained from 2010 to 2016, while for $ZYJH$ passbands there are fewer epochs, typically 2 and up to 10\,-\,12 in some fields and filters.

The variability search for this paper was focused on the Galactic plane ($|b| < 2$\degr). Aperture photometry provided by the Cambridge Astronomy Survey Unit (CASU) was used in this step. The search for CC candidates was done as in \citet{dekany2015a, dekany2015b}. The $K_S$-band time series photometry was analysed using various variability indices to perform a first selection of candidate variable objects. A periodicity analysis was carried out for this smaller sample, isolating variables in the Cepheid period range (1-100 days). 

The CC candidates in the far disk to be followed up with spectroscopic observations were obtained following the approach described in \citet{dekany2015a, dekany2015b}. Briefly, this includes computing the distances and extinctions for each Cepheid under the assumption that it belongs to each of the two classes: CC and T2C. These values are obtained from the combination of the PL relations in two near-IR bands and the corresponding observed mean magnitudes. The resulting reddening can be compared with the reddening maps from \citet[][hereafter G12]{gonzalez2012}, which provide us (assuming a reddening law) with the absolute absorption, $A_{K_S,\;(G12)}$, for bulge red clump stars, averaged over $2^\prime\times2^\prime$ tiles. Since both classes of Cepheids have very similar intrinsic $H-K_S$ colors, by assuming one type or the other we will obtain very similar extinctions but completely different distances. If the distance of the variable, under the assumption it is a T2C, is smaller than the distance to the Galactic center ($\sim8\,\rm{kpc}$) but its extinction is higher than the mean bulge value for that direction obtained from G12 maps, then the object is most probably a CC beyond the bulge. We note here that this classification was used exclusively to pre-select the sample of the best CC candidates to be observed with X-Shooter that could be found based solely on VVV photometry, and has no other use throughout the remainder of the paper\footnote{The mean magnitudes and extinctions reported in this section (hence the distances) both changed slightly from the values available when preparing the observations. The main reasons are that the available VVV photometry has improved, more accurate PL relations have been published in the literature, there have been improvements in the knowledge of the reddening law, and new near-IR templates for CCs have become available. Therefore, some of the targets would not be classified as the best candidates to be CCs by the method explained here, using the updated measurements.}.

\begin{figure}
   \centering
   \resizebox{\hsize}{!}{\includegraphics{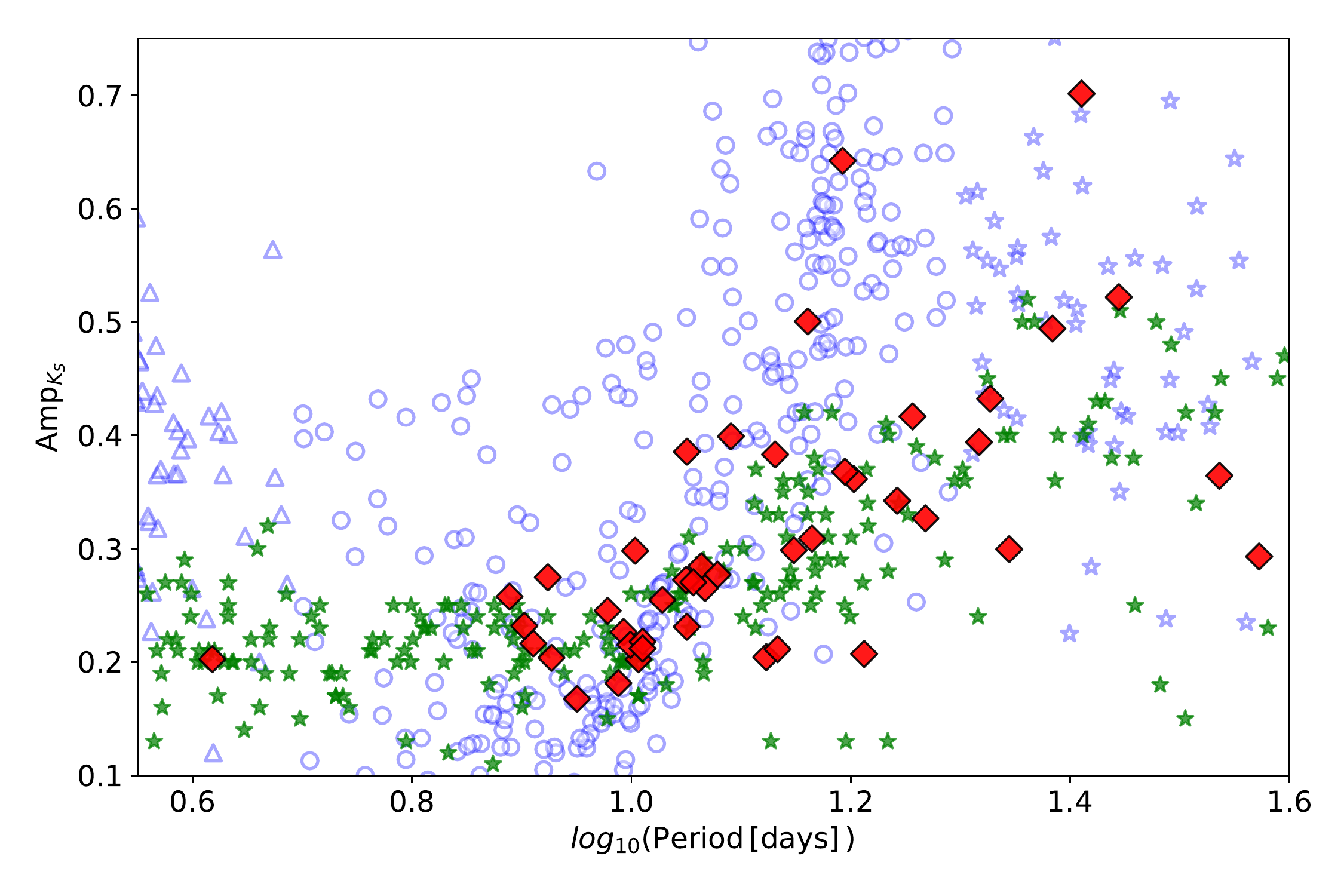}}
      \caption{Bailey diagram in $K_S$ band, showing amplitude vs. period, for our Cepheid candidate sample (red diamonds contoured in black) together with other well-known CCs (light green stars) and T2Cs (light blue open symbols). T2Cs are divided into the three main subtypes (see text). The open triangles, circles, and stars represent BL Her, W Vir, and RV Tau stars, respectively. Literature CCs come from the compilation by \citet{inno2015} and T2Cs from \citet{braga2018}. The classification of this comparison sample is based on optical data, and is therefore more accurate than those based on near-IR data.
              }
         \label{Bailey}
   \end{figure}

Figure~\ref{Bailey} shows the so-called Bailey diagram (amplitude vs. period), in the $K_S$ band, for all the Cepheids studied here together with other well-known CCs (belonging to the MW and the Magellanic Clouds) and T2Cs in the MW. It is worth mentioning here that these literature Cepheids have been accurately classified based on optical data. We note that this diagram alone does not allow us to separate the two populations clearly. The shape of the light curve in the $K$ band is also a poor indicator of the Cepheid type compared to the optical, and therefore we need to turn to spectroscopy for additional information. This figure also shows that, except for one star with $\mathrm{P}= 4.15\,{\rm days}$, all of our Cepheids are in the fundamental-mode CC period range (P $\gtrsim$ 4-6\,days). They also cover the three subtypes of T2Cs: BL Her, W Vir, and RV Tau stars. Considering the periods separating the different subtypes as defined by \citet{soszynski2011} for T2Cs in the Galactic bulge, 8 of our Cepheid candidates are in the period range $\sim$ 20-38\,days (4 with P > 25\,days), thus belonging to the RV Tau class if classified as T2Cs; 1 candidate has P$\sim$4.15\,days, corresponding to the BL Her range; and the remaining 36 have $7.7 < P < 20\,\mathrm{days}$, thus in the WVir subclass period-range.

The number of $K_{S}$-band epochs available ranges from $\sim50$ to 100, which gives a good phase coverage over the whole period range spanned by our sample. For $J$ and $H$ we have between 2-12 epochs in each band. We did not make use of the $Z$ and $Y$ photometry mainly because most of our sources are undetectable at these bands. In addition, the lack of light curve templates for $Z$ and $Y$ bands implies that their mean magnitudes would not be accurate enough. This would be even worse than for the $J$ and $H$ bands because light curve amplitudes increase with decreasing wavelength.

Due to the very high extinction towards the selected targets, some of the objects were undetectable in the $J$ band, and for this reason we used the $H$- and $K_{s}$-band mean magnitudes to simultaneously calculate the individual reddenings and distances for all our Cepheid candidates. 

From the candidate CCs obtained by this analysis, $\sim50$ were selected for spectroscopic follow-up. Throughout the paper we separate the objects into a ‘‘bulge sample’’ containing objects that are within the VVV bulge area (which covers $-10$\degr $< l < +10$\degr) and a ‘‘disk sample’’ containing objects at $l < -10$\degr.

\subsection{PSF photometry for the bulge sample}

\begin{figure}
  \resizebox{\hsize}{!}{\includegraphics{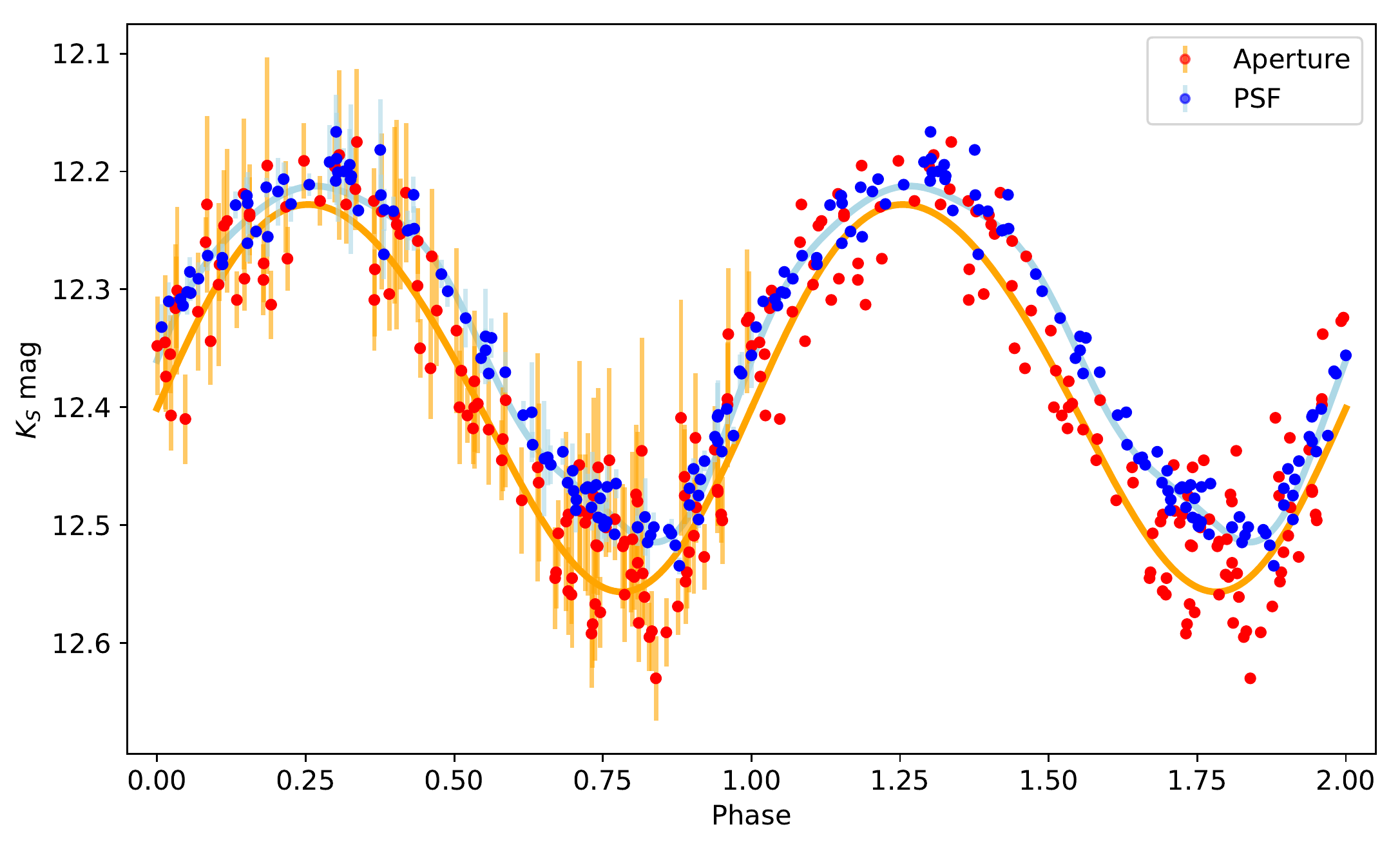}}
  \caption{$K_S$-band phased PSF (blue points) vs. aperture photometry (red points). The light blue and orange curves show the Fourier series fits (PSF and aperture, respectively).}
  \label{fig:LC}
\end{figure}

As previously mentioned, we used the aperture photometry catalogues provided by CASU to find our candidates. After the spectroscopic observations were completed, for the bulge region ($|l| \leqslant 10\degr$) we gained access to the multi-epoch point spread function (PSF) photometry performed with the DAOPHOT code \citep{stetson1987} presented in \citet{contrerasramos2017}. PSF photometry is essential in the inner region of the Galaxy approaching the Galactic plane. The stellar density increases dramatically as the disk stars are superposed with the bulge stars along the line of sight. This is the most crowded area to study on the sky. For example, Fig.~\ref{fig:LC} shows a comparison between the light curves obtained using aperture and PSF photometry for a representative sample Cepheid variable star (ID B05, see Table~\ref{tab:bulgesample}). As expected, the PSF photometry shows a significantly smaller dispersion about the fitted curve. The vertical offset between the curve and at least part of the difference in the dispersion can be explained by the calibration problem found by \citet{hajdu2019} (see next section). In what follows, we always use PSF photometry for the bulge sample.

\subsection{Recalibration of the photometry}

Recently, \citet{hajdu2019} have shown that the photometric zero-points in VVV $J$, $H$, and $K_S$ photometry present a time-varying bias. This is of great importance for the present study, given that it can affect the shapes of the light curves and significantly change the mean magnitudes obtained for the Cepheid candidates. For this reason we use recalibrated photometry in this work. For the PSF photometry used in the Galactic bulge area, we just needed to correct the zero-point of the epoch used as reference, to which the rest of the data are calibrated. Thus, the light curve shapes obtained with PSF photometry was not affected. For the disk sample we have obtained corrected $J$, $H$, and $K_S$ aperture photometry as in \citet{dekany2019}.

\subsection{Ephemerides}\label{subsection:ephemerides}

Periods and amplitudes for the targets were derived by fitting the $K_S$-band light curves with a Fourier series (order 2 to 7) using the Lomb-Scargle \citep{lomb1976,scargle1982} method from the {\tt gatspy}\footnote{\url{https://zenodo.org/record/14833}} \citep{gatspy2015} PYTHON package. As a zero-point to phase the $K_S$ light curves, we decided to use the epoch of the mean magnitude along the rising branch, $\mathrm{HJD_0}$, where “mean magnitude” means taking the average in intensity transformed into magnitude (denoted \meanks). This zero-point was suggested by \citet[][hereafter I15]{inno2015} and allowed us to adopt the near-IR light curve templates for CCs provided by these authors. This is crucial to compute accurate mean magnitudes in the $J$ band ($H$ band), where the phase coverage of our data is poor, since we have between 2 and 10 (2 and 12) epochs, 4 epochs being the most common case. 

The I15 light curve templates have some important advantages with respect to previous near-IR light curve templates \citep[e.g.,][]{soszynski2005}. First, they are based on a larger sample of Cepheids with well-sampled light curves in the $V$, $J$, $H$, and $K_S$ bands. Thus, they also have a better period sampling, with templates for ten period bins between 1 and 60\,days. Second, they have a good coverage of the Hertzsprung progression (6 < P < 16\,days), where Cepheids show a bump along their light curves whose position changes with period, which affects their shape. As can be seen in Fig.~\ref{Bailey}, the majority of the targets fall into this period range.

\begin{figure}
  \centering
  \resizebox{\hsize}{!}{\includegraphics{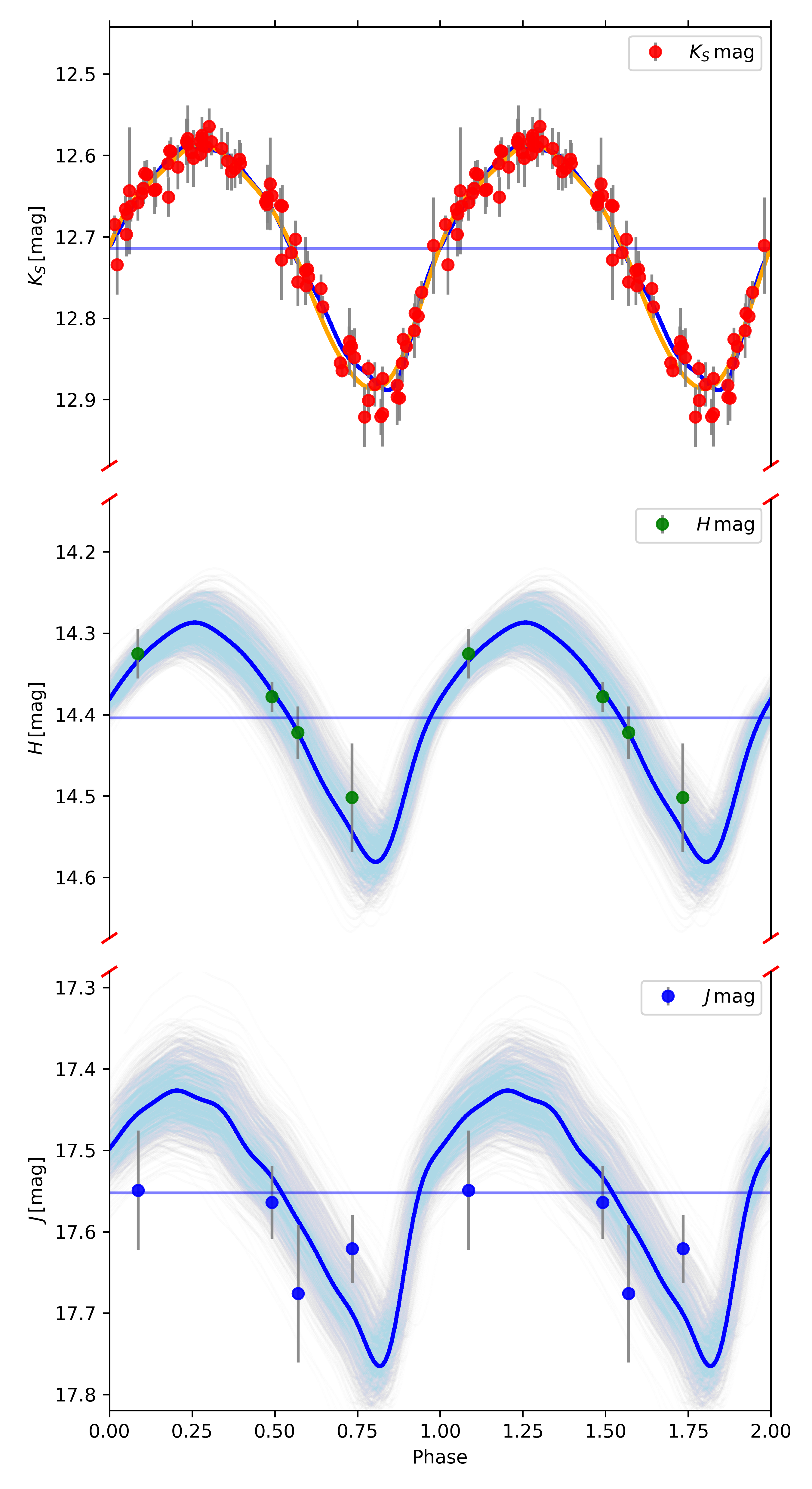}}
  \caption{Light curve fitting for B01 using I15 $H$- (middle) and $J$-band (bottom) templates. Also shown are the $K_S$-band light curve (top) together with the best-fit Fourier series (orange curve) and the corresponding I15 template (blue curve), which are almost indistinguishable from each other. The horizontal blue lines indicate the mean magnitude for each band.}
  \label{fig:HbandLC}
\end{figure}

In order to fit the templates from I15, we used the phase lags of the $J$- and $H$-band light curves with respect to the $K_S$-band light curve and the amplitude ratios (needed to estimate the $J$- and $H$-band amplitudes from our measured $\mathrm{Amp}_{K_S}$) presented in their paper. In Fig.~\ref{fig:HbandLC} we show the template fitting to the $J$-, $H$-, and $K_S$-band photometry for one of the Cepheid candidates. The uncertainties on the mean $J$ and $H$ magnitudes were calculated as the standard deviation of the mean magnitudes obtained from a thousand realizations of the template fitting procedure. In each iteration we perturbed the phase shift and amplitude of the corresponding template, according to the errors on those values reported in I15, while randomly selecting the photometric measurements in the given band from a Gaussian distribution with standard deviation equal to the photometric error. For \meanks, the uncertainty on each star was estimated as the sum in quadrature of the standard deviation of the photometric measurements around the Fourier fit and the median photometric error of the epochs used, following \citet{braga2018}.

The ephemerides and photometric information are provided in Tables~\ref{tab:bulgesample} and~\ref{tab:bulgesamplespectralinfo} for the bulge, and in Tables~\ref{tab:disksample} and~\ref{tab:disksamplespectralinfo} for the disk sample.

\begin{table*}
 \footnotesize
 \caption{Coordinates and photometric information: Bulge sample.}
 \centering
 \begin{tabular}{llccccccccr}
 \hline \hline
VVV\,ID (DR2)                                   & ID  & RA           & DEC        & \meanj               & \meanh            & \meanks  & $\mathrm{Amp}_{K_S}$ & Period        \\
                                                &     & [hh:mm:ss]   & [hh:mm:ss] & mag                   & mag               & mag      & mag                  & days     \\
\hline
J172334.10-355659.96                    & B01 & 17:23:34.10 & -35:57:00.0 & $17.469 \pm 0.067$ & $14.443 \pm 0.037$ & $12.712 \pm 0.022$ & 0.300 & 22.10092 \\
J174435.09-263209.87                    & B02 & 17:44:35.09 & -26:32:09.9 & $15.216 \pm 0.023$ & $14.004 \pm 0.034$ & $12.700 \pm 0.017$ & 0.701 & 25.72395 \\
J180052.02-232151.19                    & B03 & 18:00:52.03 & -23:21:51.2 & $18.206 \pm 0.056$ & $15.323 \pm 0.028$ & $13.892 \pm 0.021$ & 0.207 & 16.29821 \\
J175802.76-242629.86                    & B04 & 17:58:02.76 & -24:26:29.8 & $16.969 \pm 0.026$ & $13.738 \pm 0.020$ & $12.015 \pm 0.017$ & 0.285 & 11.58244 \\
J174554.44-293650.09                    & B05 & 17:45:54.44 & -29:36:50.0 & $17.174 \pm 0.040$ & $14.074 \pm 0.021$ & $12.379 \pm 0.013$ & 0.299 & 14.06155 \\
J172722.42-354036.28\tablefootmark{*}   & B06 & 17:27:22.43 & -35:40:36.3 & $16.445 \pm 0.038$ & $13.666 \pm 0.032$ & $12.187 \pm 0.015$ & 0.277 & 11.98122 \\
J180341.61-214333.99                    & B07 & 18:03:41.61 & -21:43:34.0 & $17.034 \pm 0.031$ & $13.758 \pm 0.023$ & $12.027 \pm 0.014$ & 0.265 & 11.67156 \\
J180450.03-215024.48                    & B08 & 18:04:50.03 & -21:50:24.5 & $17.367 \pm 0.041$ & $14.291 \pm 0.021$ & $12.609 \pm 0.015$ & 0.202 & 10.14914 \\
J173432.45-330512.34                    & B09 & 17:34:32.46 & -33:05:12.3 & $17.663 \pm 0.085$ & $14.201 \pm 0.038$ & $12.276 \pm 0.019$ & 0.494 & 24.19975 \\
J175903.31-241829.69                    & B10 & 17:59:03.32 & -24:18:29.7 & $16.390 \pm 0.031$ & $14.362 \pm 0.020$ & $13.433 \pm 0.016$ & 0.293 & 37.34996 \\
J180124.48-225444.63\tablefootmark{**}  & B11 & 18:01:24.49 & -22:54:44.6 & $18.489 \pm 0.086$ & $14.689 \pm 0.024$ & $12.705 \pm 0.017$ & 0.231 & 11.23291 \\
J180125.08-225428.31\tablefootmark{**}  & B12 & 18:01:25.08 & -22:54:28.3 & $18.429 \pm 0.102$ & $14.702 \pm 0.041$ & $12.673 \pm 0.016$ & 0.273 & 11.21634 \\
J175849.51-240919.80                    & B13 & 17:58:49.52 & -24:09:19.8 & $19.189 \pm 0.136$ & $15.434 \pm 0.030$ & $13.446 \pm 0.017$ & 0.417 & 18.04055 \\
J173333.20-324259.78                    & B14 & 17:33:33.21 & -32:42:59.7 & $18.553 \pm 0.076$ & $14.582 \pm 0.024$ & $12.567 \pm 0.013$ & 0.361 & 15.95089 \\
J172929.79-334126.86                    & B15 & 17:29:29.79 & -33:41:26.7 & $16.061 \pm 0.017$ & $13.928 \pm 0.020$ & $12.751 \pm 0.014$ & 0.204 & 13.27217 \\
J172643.40-345825.65                    & B16 & 17:26:43.41 & -34:58:25.6 & $20.499 \pm 0.340$ & $16.583 \pm 0.057$ & $13.850 \pm 0.018$ & 0.226 & 9.83607  \\
J180113.94-223223.72                    & B17 & 18:01:13.94 & -22:32:23.7 & $17.482 \pm 0.045$ & $14.191 \pm 0.028$ & $12.448 \pm 0.017$ & 0.270 & 11.38684 \\
J175756.63-250306.18                    & B18 & 17:57:56.63 & -25:03:06.2 & $14.498 \pm 0.015$ & $12.713 \pm 0.016$ & $11.809 \pm 0.013$ & 0.211 & 13.59042 \\
J173318.96-335616.92                    & B19 & 17:33:18.96 & -33:56:16.7 & $15.696 \pm 0.034$ & $13.763 \pm 0.029$ & $12.803 \pm 0.023$ & 0.203 & 4.14921  \\
J174232.61-304527.85                    & B20 & 17:42:32.62 & -30:45:27.8 & $16.937 \pm 0.030$ & $14.098 \pm 0.023$ & $12.661 \pm 0.018$ & 0.642 & 15.57948 \\
J174613.36-302131.54                    & B21 & 17:46:13.36 & -30:21:31.6 & $15.941 \pm 0.024$ & $13.450 \pm 0.019$ & $12.263 \pm 0.018$ & 0.383 & 13.52001 \\
J174642.13-295445.04                    & B22 & 17:46:42.14 & -29:54:45.0 & $15.360 \pm 0.031$ & $12.749 \pm 0.028$ & $11.524 \pm 0.018$ & 0.522 & 27.81618 \\
J175312.32-270054.97\tablefootmark{*}   & B23 & 17:53:12.33 & -27:00:55.1 & $16.505 \pm 0.032$ & $14.193 \pm 0.027$ & $12.988 \pm 0.017$ & 0.364 & 34.35879 \\
J175637.72-264451.58\tablefootmark{*}   & B24 & 17:56:37.73 & -26:44:51.6 & $15.833 \pm 0.037$ & $13.423 \pm 0.027$ & $12.322 \pm 0.017$ & 0.501 & 14.48111 \\
\hline                                                                                               
 \end{tabular}                                                                                       
 \label{tab:bulgesample}
 \tablefoot{
\tablefoottext{*}{ID from VVV DR1.}
\tablefoottext{**}{Invisible cluster member from \citet{dekany2015a}.}
}
 \end{table*}

\begin{table*}
 \footnotesize
 \caption{Coordinates and photometric information: Disk sample.}
 \centering
 \begin{tabular}{llccccccccr}
 \hline \hline
VVV\,ID (DR2)                           & ID  & RA           & DEC        & \meanj       & \meanh            & \meanks  & $\mathrm{Amp}_{K_S}$ & Period  \\
                                        &     & [hh:mm:ss]   & [hh:mm:ss] & mag           & mag               & mag      & mag                  & days    \\
\hline
J161044.42-523858.28                    & D01 & 16:10:44.42 & -52:38:58.3 & $14.145\pm 0.029$ & $12.968\pm 0.057$ & $12.462\pm 0.023$ & 0.342  & 17.46725 \\
J163042.97-491343.45                    & D02 & 16:30:42.97 & -49:13:43.5 & $16.502\pm 0.035$ & $14.281\pm 0.037$ & $13.294\pm 0.021$ & 0.399  & 12.32390 \\
J164612.61-470141.20                    & D03 & 16:46:12.61 & -47:01:41.4 & $15.796\pm 0.028$ & $14.290\pm 0.041$ & $13.510\pm 0.023$ & 0.245  & 9.51240  \\
J164343.50-461020.53                    & D04 & 16:43:43.50 & -46:10:20.5 &  --               & $14.665\pm 0.171$ & $12.358\pm 0.018$ & 0.368  & 15.65736 \\
J164404.55-460531.70                    & D05 & 16:44:04.55 & -46:05:31.7 &  --               & $16.071\pm 0.072$ & $13.338\pm 0.022$ & 0.215  & 9.97828  \\
J164434.14-455856.80                    & D06 & 16:44:34.14 & -45:58:56.8 &  --               & $14.707\pm 0.051$ & $12.238\pm 0.018$ & 0.432  & 21.23539 \\
J164535.11-460058.30                    & D07 & 16:45:35.11 & -46:00:58.4 & $16.229\pm 0.046$ & $14.644\pm 0.067$ & $12.893\pm 0.021$ & 0.327  & 18.53538 \\
J164651.27-454300.81                    & D08 & 16:46:51.27 & -45:43:00.8 & $17.321\pm 0.070$ & $14.151\pm 0.061$ & $12.451\pm 0.019$ & 0.255  & 10.67261 \\
J155711.88-530014.18                    & D09 & 15:57:11.89 & -53:00:14.2 &  --               & $15.457\pm 0.043$ & $13.413\pm 0.022$ & 0.204  & 8.45483  \\
J160159.86-523945.87                    & D10 & 16:01:59.87 & -52:39:45.9 & $18.899\pm 0.219$ & $15.685\pm 0.050$ & $14.119\pm 0.031$ & 0.386  & 11.23596 \\
J164009.00-460809.28                    & D11 & 16:40:09.00 & -46:08:09.4 & $17.261\pm 0.199$ & $14.383\pm 0.168$ & $12.887\pm 0.021$ & 0.216  & 8.13835  \\
J164221.76-454840.67                    & D12 & 16:42:21.77 & -45:48:40.7 &  --               & $13.798\pm 0.174$ & $12.389\pm 0.023$ & 0.218  & 10.23824 \\
J164624.29-452243.61                    & D13 & 16:46:24.30 & -45:22:43.5 & $18.143\pm 0.158$ & $14.291\pm 0.069$ & $12.476\pm 0.021$ & 0.181  & 9.72873  \\
J165835.62-423655.05                    & D14 & 16:58:35.62 & -42:36:55.1 & $18.895\pm 0.247$ & $15.268\pm 0.034$ & $13.426\pm 0.026$ & 0.232  & 7.98678  \\
J165907.62-420522.90                    & D15 & 16:59:07.63 & -42:05:22.9 & $16.425\pm 0.046$ & $13.771\pm 0.034$ & $12.403\pm 0.018$ & 0.168  & 8.92130  \\
J170759.64-402929.89                    & D16 & 17:07:59.65 & -40:29:30.0 &  --               & $15.045\pm 0.054$ & $12.836\pm 0.021$ & 0.309  & 14.60643 \\
J170339.65-395054.57                    & D17 & 17:03:39.66 & -39:50:54.6 & $15.870\pm 0.027$ & $13.788\pm 0.030$ & $12.767\pm 0.020$ & 0.212  & 10.23655 \\
J170637.80-401408.42                    & D18 & 17:06:37.81 & -40:14:08.4 & $15.592\pm 0.170$ & $13.836\pm 0.167$ & $12.892\pm 0.021$ & 0.275  & 8.39083  \\
J170640.82-400351.87                    & D19 & 17:06:40.81 & -40:03:51.9 & $17.051\pm 0.059$ & $14.749\pm 0.033$ & $13.602\pm 0.025$ & 0.298  & 10.07809 \\
J152844.79-563806.22                    & D20 & 15:28:44.79 & -56:38:06.2 & $15.205\pm 0.024$ & $13.070\pm 0.039$ & $12.014\pm 0.018$ & 0.258  & 7.74229  \\
J162500.99-491811.82\tablefootmark{*}   & D21 & 16:25:00.99 & -49:18:11.8 &  --               & $17.470\pm 0.156$ & $13.550\pm 0.024$ & 0.394  & 20.74361 \\
\hline
 \end{tabular}
 \label{tab:disksample}
 \tablefoot{
\tablefoottext{*}{ID from VVV DR1}
}
 \end{table*}

\subsection{Mid-IR data}

\begin{figure}
     \centering
     \begin{subfigure}[b]{0.24\textwidth}
         \centering
         \includegraphics[width=\textwidth]{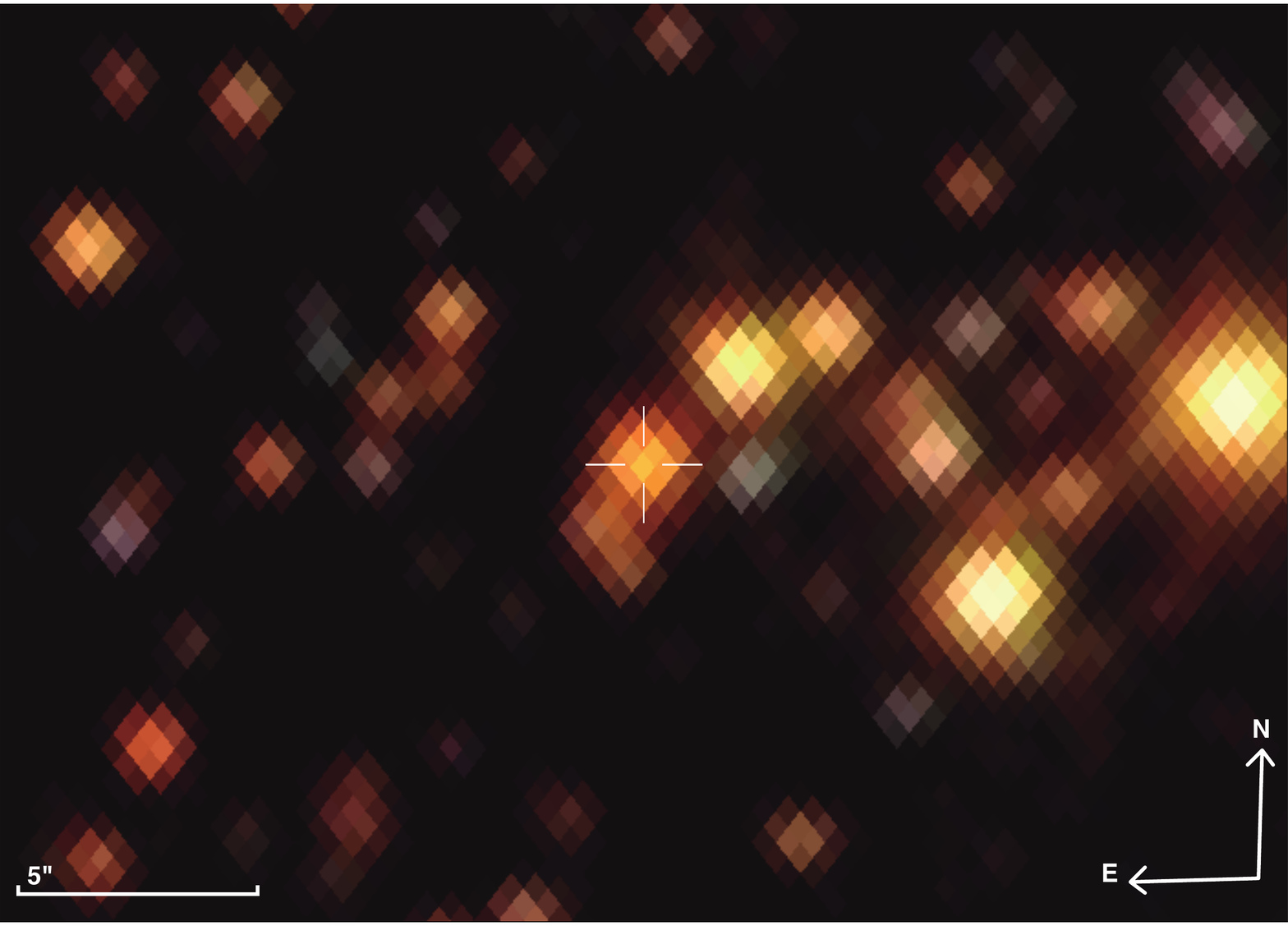}
         \label{fig:VVV}
     \end{subfigure}
     \hfill
     \begin{subfigure}[b]{0.24\textwidth}
         \centering
         \includegraphics[width=\textwidth]{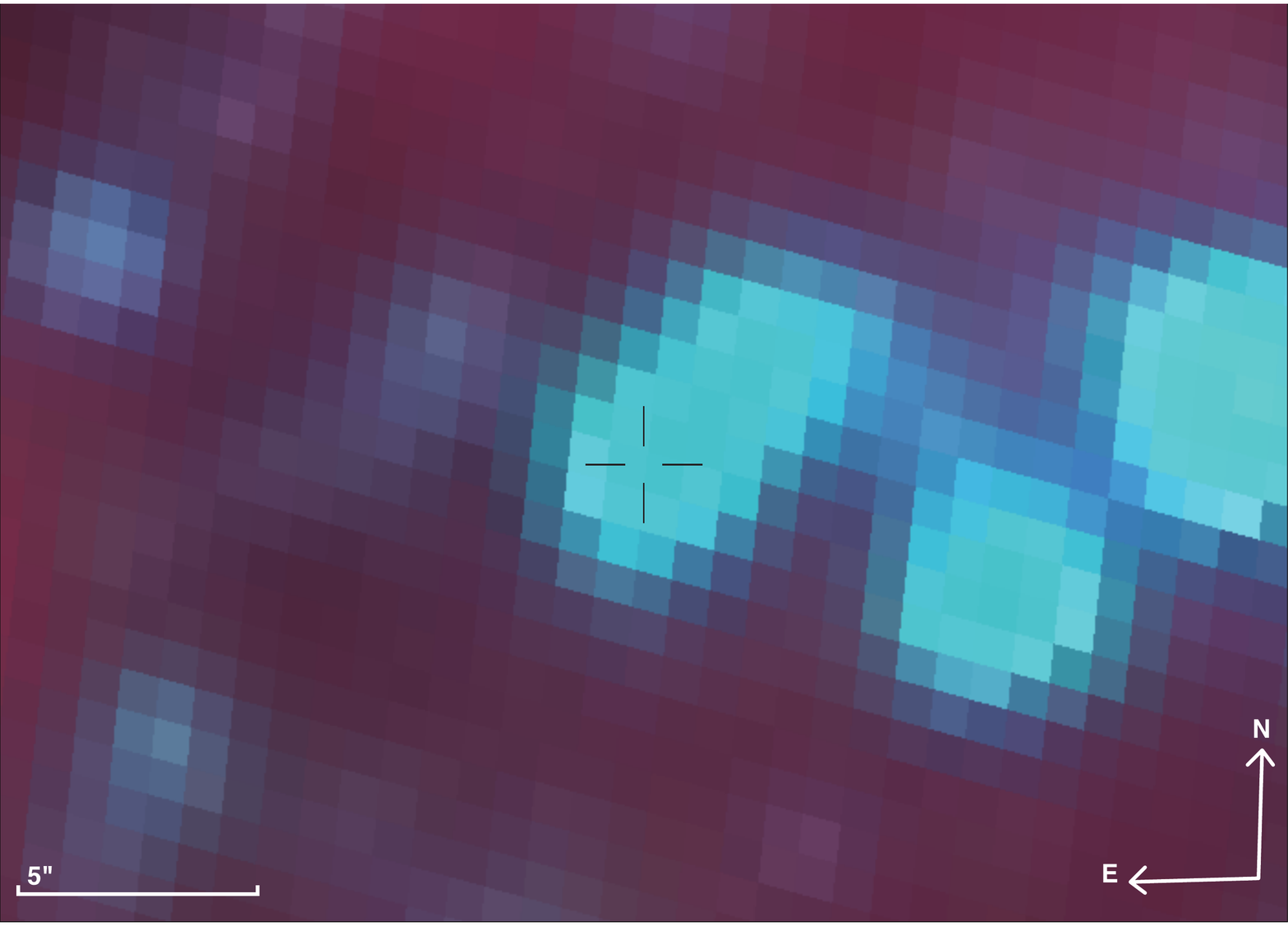}
         \label{fig:GLIMPSE}
     \end{subfigure}
        \caption{Small field around B01 illustrating the effect of crowding and the need for a high spatial resolution in the studied region of the Galaxy. Left: VVV color image. Right: GLIMPSE color image for the same region. North is up and east is to the left. A $5\arcsec$ scale is indicated in the lower left corner of the image.
                }
        \label{fig:glimpse}
\end{figure}

We also retrieved mid-IR magnitudes from the Spitzer Galactic Legacy Infrared Mid-Plane Survey Extraordinaire (GLIMPSE, I/II/3D/DEEP) \citep{benjamin2003,churchwell2009} survey in the four IRAC bands available: $[3.6]$, $[4.5]$, $[5.8]$, and $[8.0]$ $\mu m$. We found a match for all the Cepheid candidates except one (D03).

When calculating the distances from the combination of $K_S$ and mid-IR photometry, we found that for some of the stars we had obtained shorter distances than those derived from the VVV bands alone. Most likely these sources do not have reliable mid-IR magnitudes; in other words because of the severe crowding present in the bulge and the lower spatial resolution of GLIMPSE data, their mid-IR magnitudes are brighter than they should be. Figure~\ref{fig:glimpse} shows the VVV and GLIMPSE color images for B01. From the VVV image we see that there is a red source at $\sim1.2\arcsec$ to the southeast of this Cepheid candidate that cannot be resolved by Spitzer (spatial resolution of $\sim1.4\arcsec\rm{-}\,2\arcsec$). This explains why we measure a larger distance and lower extinction for this source when calculating them from the combination of the $K_S$-band and $[3.6]$ magnitudes compared to the values obtained using the VVV $H$ and $K_S$ bands. For this reason, we do not use GLIMPSE photometry in this work. Care has to be taken when using mid-IR photometry in the inner regions of the Galaxy.

\subsection{Distance determination and the extinction law}\label{subsection:reddening_law}

In order to determine the distance and reddening for each object, we obtained their absolute magnitudes using the near-IR PL relations derived by \citet{macri2015} for CCs and by \citet{bhardwaj2017} for T2Cs. The PL relations were transformed from the 2MASS to VISTA photometric system using the empirical transformations presented by \citet{gonzalez-fernandez2018}. The distance modulus to the Large Magellanic Cloud recently derived by \citet{pietrzynski2019}, $\mu_{0,LMC} = 18.477$, was used to calibrate the zero points. As a result, we obtain the following PL relations for CCs:

\begin{equation} \label{eq:1}
M_\mathrm{J}  = -3.159 \times \left(log_{10}\left(P\right) - 1\right) - 5.263,
\end{equation}
\begin{equation} \label{eq:2}
M_\mathrm{H} = -3.186 \times \left(log_{10}\left(P\right) - 1\right) - 5.623,
\end{equation}
\begin{equation} \label{eq:3}
M_\mathrm{K_S} = -3.248 \times \left(log_{10}\left(P\right) - 1\right) - 5.704;
\end{equation}
and for T2Cs, if $log_{10}(P)<1.3$,
\begin{equation} \label{eq:4}
M_\mathrm{J} = -2.066 \times \left(log_{10}\left(P\right) - 1\right) - 2.929,
\end{equation}
\begin{equation} \label{eq:5}
M_\mathrm{H} = -2.199 \times \left(log_{10}\left(P\right) - 1\right) - 3.328,
\end{equation}
\begin{equation} \label{eq:6}
M_\mathrm{K_S} = -2.233 \times \left(log_{10}\left(P\right) - 1\right) - 3.410;
\end{equation}
otherwise,
\begin{equation} \label{eq:7}
M_\mathrm{J} = -2.247 \times \left(log_{10}\left(P\right) - 1\right) - 3.348,
\end{equation}
\begin{equation} \label{eq:8}
M_\mathrm{H} = -2.298 \times \left(log_{10}\left(P\right) - 1\right) - 3.718,
\end{equation}
\begin{equation} \label{eq:9}
M_\mathrm{K_S} = -2.173 \times \left(log_{10}\left(P\right) - 1\right) - 3.826
.\end{equation}

Knowing the extinction law is critical at this point. We need to determine, for each star, the total extinction in $K_S$ ($A_{K_S}$) from its observed color excess; this means that we need an accurate prediction of the total-to-selective extinction ratio for the combination of filters used. As discussed in \citet{matsunaga2016}, differences in the adopted extinction law have a direct impact on the determination of distances. This is especially important for the low Galactic latitude regions studied in this paper, considering the high reddenings of our objects ($E(H-K_S)>0.8 \, \rm{mag}$), and thus represents the main source of systematic uncertainty in the derived distances. To take this problem into account, \citet{tanioka2017} and \citet{inno2019} used extreme values of the selective-to-total extinction ratios present in the literature to estimate the systematic uncertainty on their derived distances. It is worth noting here that the selective-to-total extinction ratio $A_{K_S}/E(H-K_S)$ values available in the literature for the inner regions of the Galaxy span a wide range, going from 1.10 \citep{alonso-garcia2017} to 1.61 \citep{nishiyama2009}, while for $A_{K_S}/E(J-K_S)$ they are constrained to 0.428-0.528, with the lower and upper limit coming from these same two references. Therefore, the errors derived in this way will be very large.

In this work, we decided to take a different approach. We selected the variables that we classified as CCs (see Sect.~\ref{subsection:CCs}), and used them to study the extinction law towards these highly reddened regions of our Galaxy. This can be done thanks to the very accurate PL relations followed by CCs, particularly in the near-IR, where their typical dispersions are less than $\sim 0.10 \, \rm{mag}$ \citep{inno2016}. From the PL relations, we can derive accurate intrinsic colors that, when subtracted from the observed colors for each sample star, give us their color excesses at the different combinations of observed bands. Thus, we can determine their color excess ratios and calculate the near-IR extinction law.

To make a first selection of CCs in our sample, we performed a preliminary classification following the same procedure described in Sect.~\ref{subsection:classificationtype}. At this point the distances to each object (needed for the classification process) were calculated using the selective-to-total extinction ratio $A_{K_S}/E(H-K_S)=1.44$ reported by \citet{nishiyama2006}, a value between those derived in the two works mentioned above. Hence, we obtained a clean sample of CCs. Considering that our goal was to study the near-IR extinction law, we used only those with reliable mean magnitudes in the three bands (15 stars). It is worth mentioning that this classification did not change when we adopted the final selective-to-total extinction ratios derived in what follows.

In order to increase the size of the sample of CCs with $J$, $H$, and $K_S$ magnitudes, we added the stars present in the Optical Gravitational Lensing Experiment (OGLE) collection of Galactic Cepheids \citep{udalski2018} that were also observed with VVV. We take the photometric information for these stars from \citet{dekany2019}, that did a cross-match between the OGLE catalogue and their own Cepheid sample found using VVV photometry, and provide their periods and mean $J$, $H$, and $K_S$ magnitudes. These are {bona fide} CCs (considering that they were classified based on well-sampled light curves in the optical) that not only allow us to increase the sample size, but also have lower reddening values than the CCs studied in the present work, and thus they widen the range of color excesses spanned by the sample of CCs used to study the near-IR extinction law. As a trade-off for the lower color excesses, many of the OGLE CCs have \meanks$\,<12\,\rm{mag,}$ and are thus saturated in VVV photometry. For this reason, we could only add five OGLE CCs (OGLE-GD-CEP-0884, OGLE-GD-CEP-0973, OGLE-GD-CEP-1003, OGLE-BLG-CEP-065, OGLE-BLG-CEP-089) and ended up with a sample of 20 {bona fide} CCs with accurate near-IR photometry. We obtained the following mean color excess ratios
   \begin{displaymath}
      1) \frac{E(J-K_S)}{E(H-K_S)} = 2.812\pm0.028  \, , 
   \end{displaymath}
   
   \begin{displaymath}
      2) \frac{E(J-H)}{E(J-K_S)} = 0.6441\pm0.0035 \, ,
   \end{displaymath} 
where the errors were estimated taking into account the uncertainties of the photometry and the PL relations.

Assuming that the extinction law in the near-IR ($JHK_{s}$) can be approximated by a power law of the form $A_{\lambda}\propto\lambda^{-\alpha}$, from the color excess ratios we can measure the power-law index $\alpha$. The effective wavelengths ($\lambda_{eff}$) for the VISTA $J$, $H$, and $K_S$ filters are $1.254\,\mathrm{\mu m}$, $1.646\,\mathrm{\mu m,}$ and $2.149\,\mathrm{\mu m}$, respectively.

From 1) and 2) we get $\alpha = 2.13\pm0.06$. Based on the obtained power-law index value, we can estimate the absolute extinction ratios as $A_{\lambda}/A_{K_S}=(\lambda_{K_S}/\lambda)^\alpha$. For $\alpha=2.13\pm0.06$, $A_{J}/A_{K_S}=3.15\pm0.10$ and $A_{H}/A_{K_S}=1.76\pm0.03$. The selective-to-total extinction ratios are
   \begin{displaymath}
      \frac{A_{K_S}}{E(J-K_S)} = 0.465\pm0.022  \, , \; \frac{A_{K_S}}{E(H-K_S)} = 1.308\pm0.050.
   \end{displaymath}

The value of $A_{K_S}/E(J-K_S)$ obtained with this sample of CCs is in good agreement with recent studies. \citet{minniti2018} estimate $A_{K_S}/E(J-K_S)=0.484\pm0.040$ using the red clump method \citep{nishiyama2006} on VVV data for a low extinction window at $(l,b)=(-12.6$\degr$,-0.4$\degr$),$ and \citet{majaess2016} measure a value of 0.49 using different populations of pulsating variable stars. Recently, \citet{dekany2019} obtained $A_{K_S}/E(J-K_S)=0.528\pm0.019$ using a sample of more than 1200 Cepheids found with VVV data.

We used the selective-to-total extinction ratios derived in this section to calculate the $A_{K_S}$ for the sample of Cepheids. Distances and extinctions were calculated for each Cepheid candidate under the assumption that it is both a CC and a T2C. Whenever the $J$-band mean magnitude was available and its uncertainty was less than $0.20\,\rm{mag}$, we calculated for each star a pair of distances from the combination of both the $J$ $\&$ $K_S$ and $H$ $\&$ $K_S$ PL relations, with their associated errors. In these cases, the final heliocentric distances presented in Tables~\ref{tab:bulgesamplespectralinfo} and ~\ref{tab:disksamplespectralinfo} were computed as the weighted average using both distances. The individual distance errors were obtained from Monte Carlo simulations, taking into account the uncertainty on the mean $J$, $H$, and $K_S$ magnitudes, both the intrinsic dispersion and the uncertainties on the parameters of PL relations used (see \citealp{macri2015} and \citealp{bhardwaj2017} for CCs and T2Cs, respectively)
and the errors on the calculated selective-to-total extinction ratios.

\section{Spectroscopic observations}\label{section:spectroscopy}

The spectroscopic follow-up observations for the selected sample of 45 Cepheid candidates were carried out in service mode, with the X-Shooter spectrograph \citep{vernet2011} located at the ESO Very Large Telescope, in the second semester of 2015 (ESO program ID 095.B-0444(A), PI: Dékány). This instrument provides full spectral coverage from the UV to the near-IR ($3100-24800$ \AA) at medium resolution. We used the pipeline-reduced Internal Data Products provided by ESO.

As previously stated, the targets are highly obscured objects. For this reason, even if the range of temperatures covered by CCs is $\sim4700-6500\mathrm{K}$, these stars can only be detected with the near-IR arm of X-Shooter (providing a wavelength coverage from 1024 to 2480 nm). Single-epoch spectra were obtained at random phase. The exposure times were set in order to achieve a signal-to-noise ratio (S/N) $\geqslant 50$ for the $K$-band spectral range, where the candidate Cepheids have higher fluxes. Typical exposure times were 7-50 min. The observations were taken with seeing better than $1\arcsec\,$in $V$ band, thus better than $0.8\arcsec$ in the $K_S$ band. The slit width was set to $0.6\arcsec$ and the corresponding resolution in the near-IR was $R\sim8000$. Given that the targets are pulsating variable stars, their measured radial velocities (RVs) need to be corrected by pulsational effects. This correction will be presented in Sect.~\ref{section:kinematics}. Three objects have been observed at more than one epoch: B05 and D05 have two observations, and B20 has three (see Tables~\ref{tab:bulgesamplespectralinfo} and \ref{tab:disksamplespectralinfo}.

\begin{table*}
 \footnotesize
 \caption{Spectroscopic observations, ephemerides, distances, and extinctions: Bulge sample.}
 \centering
 \begin{tabular}{lcccccccccc}
 \hline \hline
 ID &   $l$    &   $b$    & <$K_S$> & $\mathrm{S/N}_{K_S}$  & $\mathrm{HJD_0}$\tablefootmark{a} & $\mathrm{HJD_{obs}}$\tablefootmark{a} & $ \Phi_\mathrm{obs}$ & $\mathrm{d_{T2C}}$ & $\mathrm{d_{CC}}$ & $A_{K_S,\;\mathrm{CC}}$ \\
    &  [\degr]  &  [\degr]  &   mag   &                       &    days & days        &                      & kpc                & kpc               & mag                     \\
\hline
B01 & -8.41212 & 0.10398  & 12.71 & 43.2  & 57255.494284 & 57251.616021  & 0.825 & $11.00 \pm 1.26$ & $31.46 \pm 1.34$ & 2.06  \\
B02 & 1.92811  & 1.44779  & 12.70 & 60.4  & 57273.964842 & 57235.720726  & 0.513 & $16.66 \pm 1.87$ & $50.19 \pm 1.70$ & 1.11  \\
B03 & 6.53638  & -0.11529 & 13.89 & 33.3  & 57282.465888 & 57235.628966  & 0.126 & $15.95 \pm 1.11$ & $50.83 \pm 1.94$ & 1.77  \\
B04 & 5.28099  & -0.0926  & 12.02 & 47.0  & 57283.374023 & 57273.571803  & 0.154 & $4.92  \pm 0.35$ & $14.59 \pm 0.58$ & 2.12  \\
B05 & -0.54593 & -0.40642 & 12.38 & 58.9  & 57269.109205 & 57282.528352  & 0.954 & $6.51  \pm 0.46$ & $20.11 \pm 0.78$ & 2.06  \\
B05 & -0.54593 & -0.40642 & 12.38 & 57.6  & 57269.109205 & 57293.546495  & 0.738 & $6.51  \pm 0.46$ & $20.11 \pm 0.78$ & 2.06  \\
B06 & -7.75337 & -0.38173 & 12.19 & 48.2  & 57249.738471 & 57233.749423  & 0.665 & $6.27  \pm 0.43$ & $18.74 \pm 0.69$ & 1.80  \\
B07 & 8.28308  & 0.12610  & 12.03 & 71.3  & 57274.584841 & 57278.605523  & 0.344 & $4.92  \pm 0.35$ & $14.64 \pm 0.58$ & 2.14  \\
B08 & 8.31370  & -0.16051 & 12.61 & 55.8  & 57276.216131 & 57235.745653  & 0.012 & $6.28  \pm 0.44$ & $18.16 \pm 0.70$ & 2.05  \\
B09 & -4.77693 & -0.19486 & 12.28 & 99.5  & 57256.446017 & 57283.529906  & 0.119 & $8.23  \pm 0.96$ & $24.08 \pm 1.10$ & 2.33  \\
B10 & 5.51131  & -0.22519 & 13.43 & 40.2  & 57303.134583 & 57235.674511  & 0.194 & $29.28 \pm 3.25$ & $95.59 \pm 3.27$ & 1.11  \\
B11 & 6.99048  & 0.00053  & 12.71 & 55.5  & 57274.235472 & 57282.546168  & 0.740 & $5.63  \pm 0.41$ & $16.63 \pm 0.74$ & 2.48  \\
B12 & 6.99554  & 0.00081  & 12.67 & 61.2  & 57279.222133 & 57290.544655  & 0.009 & $5.48  \pm 0.41$ & $16.18 \pm 0.75$ & 2.51  \\
B13 & 5.61755  & -0.10356 & 13.45 & 22.9  & 57277.501613 & 57290.566667  & 0.724 & $9.75  \pm 1.00$ & $31.98 \pm 1.96$ & 2.47  \\
B14 & -4.57864 & 0.18020  & 12.57 & 40.7  & 57260.714748 & 57293.496418  & 0.055 & $6.02  \pm 0.44$ & $19.12 \pm 0.85$ & 2.54  \\
B15 & -5.85859 & 0.35613  & 12.75 & 46.3  & 57264.368617 & 57290.497841  & 0.969 & $10.32 \pm 0.69$ & $31.51 \pm 1.00$ & 1.38  \\
B16 & -7.24358 & 0.12015  & 13.85 & 35.2  & 57247.699417 & 57255.554578  & 0.799 & $5.70  \pm 0.66$ & $16.43 \pm 1.31$ & 3.47  \\
B17 & 7.29401  & 0.22020  & 12.45 & 63.5  & 57283.780522 & 57290.526367  & 0.592 & $5.87  \pm 0.42$ & $17.37 \pm 0.70$ & 2.15  \\
B18 & 4.74108  & -0.37774 & 11.81 & 56.1  & 57278.133703 & 57273.561221  & 0.664 & $7.84  \pm 0.51$ & $24.08 \pm 0.70$ & 1.05  \\
B19 & -5.63033 & -0.44288 & 12.80 & 40.4  & 57255.109399 & 57216.673404  & 0.737 & $6.92  \pm 0.46$ & $16.71 \pm 0.58$ & 1.17  \\
B20 & -1.90105 & -0.38596 & 12.66 & 28.9  & 57254.513790 & 57251.641793  & 0.816 & $8.89  \pm 0.61$ & $28.06 \pm 1.03$ & 1.77  \\
B20 & -1.90105 & -0.38596 & 12.66 & 30.7  & 57254.513790 & 57252.587705  & 0.876 & $8.89  \pm 0.61$ & $28.06 \pm 1.03$ & 1.77  \\
B20 & -1.90105 & -0.38596 & 12.66 & 51.6  & 57254.513790 & 57293.527548  & 0.504 & $8.89  \pm 0.61$ & $28.06 \pm 1.03$ & 1.77  \\
B21 & -1.14665 & -0.85219 & 12.26 & 46.8  & 57281.818089 & 57278.591086  & 0.761 & $7.97  \pm 0.53$ & $24.44 \pm 0.82$ & 1.47  \\
B22 & -0.71138 & -0.70902 & 11.52 & 70.6  & 57259.668546 & 57273.550491  & 0.499 & $8.94  \pm 1.01$ & $27.20 \pm 1.00$ & 1.51  \\
B23 & 2.50866  & -0.44817 & 12.99 & 43.3  & 57273.148312 & 57235.701716  & 0.910 & $20.05 \pm 2.24$ & $64.04 \pm 2.33$ & 1.42  \\
B24 & 3.12520  & -0.97188 & 12.32 & 42.7  & 57282.164197 & 57247.654525  & 0.617 & $8.83  \pm 0.59$ & $27.44 \pm 0.92$ & 1.37  \\
\hline
 \end{tabular}
 \label{tab:bulgesamplespectralinfo}
 \tablefoot{
 \tablefoottext{a}{HJD - 2400000.0}
 }
 \end{table*}

\begin{table*}
 \footnotesize
 \caption{Spectroscopic observations, ephemerides, distances, and extinctions: Disk sample.}
 \centering
 \begin{tabular}{lcccccccccc}
 \hline \hline
 ID &   $l$    &   $b$    & <$K_S$> & $\mathrm{S/N}_{K_S}$ & $\mathrm{HJD_0}$\tablefootmark{a} & $\mathrm{HJD_{obs}}$\tablefootmark{a} & $ \Phi_\mathrm{obs}$ & $\mathrm{d_{T2C}}$ & $\mathrm{d_{CC}}$ & $A_{K_S,\;\mathrm{CC}}$ \\
    &  [\degr]  &  [\degr]  &   mag   &                      &    days   & days          &                      & kpc                & kpc               & mag                     \\
\hline

D01 & 330.55210 & -0.81085 & 12.46 & 65.8  & 57229.871246 & 57230.630527 & 0.043 & $14.95 \pm 1.66$ & $48.14 \pm 1.55$ & 0.56   \\
D02 & 335.19359 & -0.59976 & 13.29 & 39.2  & 57227.062075 & 57248.568514 & 0.745 & $13.67 \pm 1.52$ & $41.18 \pm 1.40$ & 1.24   \\
D03 & 338.55770 & -1.05480 & 13.51 & 37.3  & 57229.103958 & 57230.656323 & 0.163 & $15.80 \pm 1.05$ & $45.19 \pm 1.45$ & 0.88   \\
D04 & 338.92785 & -0.17303 & 12.36 & 58.4  & 57237.910758 & 57246.601838 & 0.555 & $4.58  \pm 0.68$ & $14.57 \pm 1.78$ & 2.89   \\
D05 & 339.02829 & -0.16642 & 13.34 & 30.5  & 57230.261826 & 57233.732321 & 0.348 & $4.53  \pm 0.54$ & $13.11 \pm 1.11$ & 3.47   \\
D05 & 339.02829 & -0.16642 & 13.34 & 34.1  & 57230.261826 & 57276.510960 & 0.635 & $4.53  \pm 0.54$ & $13.11 \pm 1.11$ & 3.47   \\
D06 & 339.16724 & -0.15944 & 12.24 & 66.0  & 57236.224003 & 57246.618517 & 0.489 & $5.33  \pm 0.59$ & $15.32 \pm 1.14$ & 3.10   \\
D07 & 339.25675 & -0.31525 & 12.89 & 52.7  & 57238.080973 & 57246.642958 & 0.462 & $8.84  \pm 0.95$ & $29.13 \pm 2.01$ & 2.16   \\
D08 & 339.62784 & -0.28889 & 12.45 & 43.3  & 57228.512506 & 57247.551236 & 0.784 & $5.88  \pm 0.43$ & $17.20 \pm 0.75$ & 2.08   \\
D09 & 328.78550 & 0.27945  & 13.41 & 42.7  & 57219.399935 & 57230.503896 & 0.313 & $6.58  \pm 1.18$ & $18.40 \pm 1.21$ & 2.57   \\
D10 & 329.55777 & 0.06621  & 14.12 & 29.1  & 57212.978009 & 57230.544132 & 0.563 & $13.81 \pm 2.49$ & $41.02 \pm 2.66$ & 1.94   \\
D11 & 338.54691 & 0.31637  & 12.89 & 37.2  & 57226.460783 & 57246.663001 & 0.482 & $7.14  \pm 0.61$ & $19.77 \pm 1.17$ & 1.84   \\
D12 & 339.04464 & 0.24256  & 12.39 & 39.5  & 57231.744140 & 57247.562923 & 0.545 & $6.57  \pm 0.95$ & $19.15 \pm 2.23$ & 1.73   \\
D13 & 339.83419 & -0.00972 & 12.48 & 55.7  & 57233.004060 & 57273.496588 & 0.162 & $5.05  \pm 0.39$ & $14.51 \pm 0.73$ & 2.34   \\
D14 & 343.36043 & 0.05028  & 13.43 & 36.9  & 57234.721739 & 57233.670945 & 0.868 & $7.28  \pm 0.76$ & $20.12 \pm 1.25$ & 2.31   \\
D15 & 343.83398 & 0.29871  & 12.40 & 58.3  & 57225.679017 & 57273.510887 & 0.362 & $6.41  \pm 0.44$ & $18.06 \pm 0.66$ & 1.68   \\
D16 & 346.11571 & -0.04920 & 12.84 & 42.8  & 57244.818008 & 57273.527736 & 0.966 & $5.86  \pm 0.64$ & $18.38 \pm 1.30$ & 2.77   \\
D17 & 346.12866 & 0.99896  & 12.77 & 38.1  & 57237.571245 & 57278.567895 & 0.005 & $9.90  \pm 0.66$ & $28.66 \pm 0.94$ & 1.23   \\
D18 & 346.16441 & 0.31225  & 12.89 & 41.1  & 57235.244495 & 57282.503178 & 0.632 & $10.29 \pm 0.83$ & $28.78 \pm 1.44$ & 1.06   \\
D19 & 346.30697 & 0.40765  & 13.60 & 35.4  & 57233.471895 & 57233.707684 & 0.023 & $13.39 \pm 0.92$ & $38.66 \pm 1.38$ & 1.40   \\
D20 & 323.33013 & -0.12655 & 12.01 & 122.2 & 57204.570104 & 57230.585699 & 0.360 & $6.03  \pm 0.41$ & $16.52 \pm 0.55$ & 1.28   \\
D21 & 334.49553 & 0.01898  & 13.55 & 35.7  & 57242.315913 & 57248.523431 & 0.299 & $4.04  \pm 0.65$ & $11.53 \pm 1.58$ & 4.99   \\
\hline
 \end{tabular}
 \label{tab:disksamplespectralinfo}
 \tablefoot{
 \tablefoottext{a}{HJD - 2400000.0}
 }
 \end{table*}

\subsection{Telluric correction}

\begin{figure*}
\sidecaption
  \includegraphics[width=12cm]{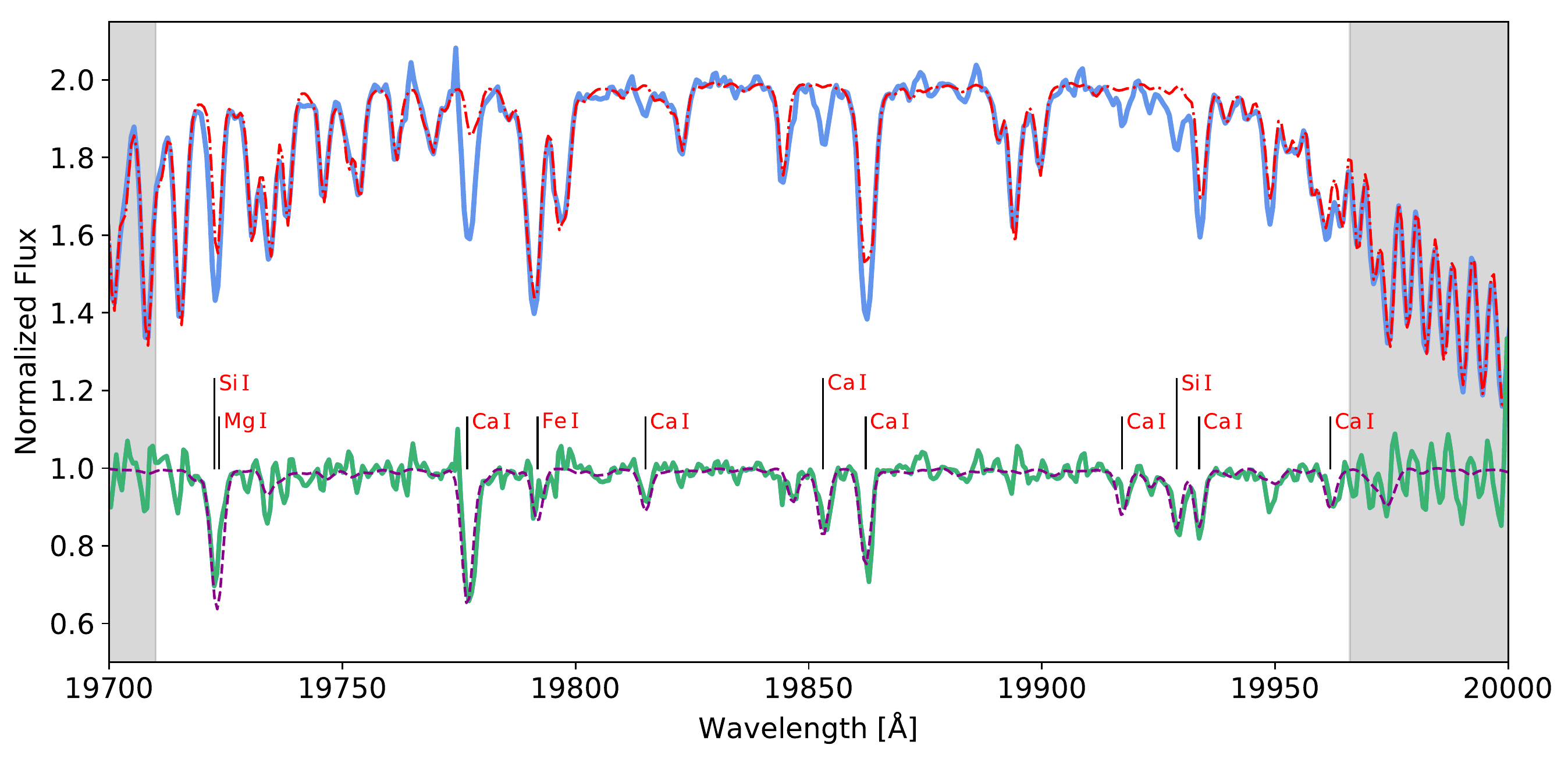}
     \caption{Section of the X-Shooter spectrum for one of the observed Cepheid candidates (B06). Top: Observed spectrum (light blue) with the best-fit telluric model (red  dot-dashed line). This spectral region is particularly affected by telluric absorption. Bottom: Final telluric-corrected spectrum (green) together with its corresponding best-fit synthetic spectrum (magenta  dashed line). The most prominent absorption lines are marked and labeled. The gray shaded regions were excluded from the spectral fitting procedure due to the large residuals in the telluric correction process of these heavily affected spectral regions (see Sect.~\ref{section:classification} for more details about the atmospheric parameter determination).}
     \label{fig:spectrum}
\end{figure*}

The spectra are strongly affected by the telluric absorption features caused by the Earth's atmosphere, as can be seen in Fig.~\ref{fig:spectrum}. This effect is particularly important in the near-IR, where several molecules (mainly water vapor, but also $\mathrm{O_2}$, CO$_{2}$, CH$_4$, CO) absorb the light coming from the targets before reaching the instrument. In order to correct for this, we used a tool available within the ESO Sky Correction Tools, called Molecfit \citep{smette2015}, that has been extensively tested on X-Shooter spectra \citep{kausch2015}.  

The recommended values for the parameter file for the near-IR arm of X-Shooter were used for this procedure. The only molecule allowed to vary for the fit was water vapor. The abundance of CO$_2$ was fixed to 1.07 in order to account for the increase in the global concentration of this molecule in the atmosphere, as done by \citet{kausch2015}. We selected the spectral windows to fit the atmospheric model based on the S/N of each spectrum. Given the typically large extinctions that our objects undergo, most of the target spectra do not have flux in the $J$ band and in some extreme cases not even in the $H$ band where the flux was low and the S/N poor. Figure~\ref{fig:spectrum} shows the best synthetic transmission spectrum fit for a representative target, and the corrected spectrum after the subtraction of the atmospheric absorption features. In spite of the seriousness of the telluric absorption in some spectral regions, as the one shown in Fig.~\ref{fig:spectrum}, this was successfully removed in all cases, and the results are not influenced by this contamination. It is also worth mentioning that the region we are showing in this figure is particularly strongly affected when compared with the typical regions used in the determination of the atmospheric parameters.

\section{Kinematics and radial velocities}\label{section:kinematics}

The individual RVs were measured by means of the cross-correlation technique and using a sample of synthetic spectra covering the typical range of atmospheric parameters of CCs and T2Cs. 

\begin{figure}
        \centering
        \includegraphics[width=\hsize]{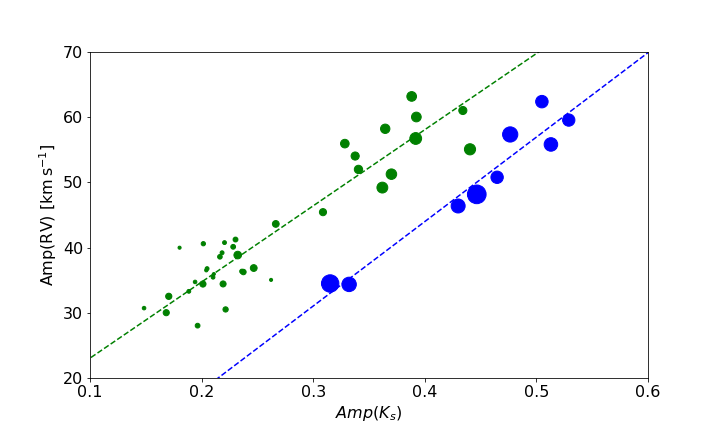}
                \caption{Relation between $\mathrm{Amp}_{K}$ and $\mathrm{Amp_{V_r}}$ for the template Cepheids. We find two relations, for periods $P<22$ days (green dashed line) or $P>22$ days (blue dashed line). The size of the points increases with the period.}
                \label{fig:ampksrv}
        \end{figure}

When considering pulsating variable stars the observed heliocentric RVs ($V_\mathrm{r,HC}(\Phi_\mathrm{obs})$) include the component associated with the systemic RV ($V_{\gamma}$) and the radial velocity component due to its pulsation ($V_{\rm r,puls}$). For Cepheids, the pulsational velocity amplitudes range from $\sim30$\,\kms{} to more than 60\,\kms. This is shown in Fig.~\ref{fig:ampksrv} where we plotted the radial velocity amplitudes, $\mathrm{Amp_{V_{r,puls}}}$, as a function of $\mathrm{Amp}_{K}$ for a sample of {bona fide} CCs compiled by M.A.T.~Groenewegen \citep[see][hereafter G13]{groenewegen2013}. Thus, given that we have single-epoch observations, we need to estimate the pulsational velocity $V_\mathrm{r,puls}(\Phi_\mathrm{obs})$ at the observed phase (denoted $\Phi_\mathrm{obs}$) and subtract it from the measured velocity $V_\mathrm{r,HC}(\Phi_\mathrm{obs})$.

To do this correction, we used the sample of $\sim$50 {bona fide} CCs provided by M.A.T.~Groenewegen (private communication) with $K$-band light curves together with their RV curves. We used this sample to construct templates of the $V_\mathrm{r,puls}$ curve. These templates were built selecting the two to six closest CCs in the period versus amplitude diagram to each of the objects studied in this work.

As done for our objects (see Sect. \ref{subsection:ephemerides}), we phased the individual\defcitealias{groenewegen2013}{G13}\citetalias{groenewegen2013} $K$-band light curves and RV curves taking the epoch of \meanks~along the rising branch as the zero point. We normalized the phased RV curves, setting the mean equal to zero and the total amplitude equal to one. Then we retrieved the value of the normalized RV template at the phase of the X-Shooter observation. This value was multiplied by the predicted velocity amplitude for that particular object, obtained from its observed $\mathrm{Amp}_{K_S}$ and the $\mathrm{Amp}_{K}$ versus $\mathrm{Amp_{V_r}}$ relations shown in Fig.~\ref{fig:ampksrv}. We found that they follow two different relations, for periods lower and higher than 22 days. The resulting velocity is the radial velocity associated to the pulsation at the phase of the spectroscopic observation, $V_\mathrm{r,puls}(\Phi_\mathrm{obs})$. We subtract this value from $V_\mathrm{r,HC}(\Phi_\mathrm{obs})$ and obtain an estimate of $V_{\gamma}$.

The measured $V_\mathrm{r,HC}(\Phi_\mathrm{obs})$, and $V_{\gamma}$ for the bulge and disk samples are presented in Table~\ref{tab:bulgesampleparameters} and Table~\ref{tab:disksampleparameters}, respectively. It is important to keep in mind that the correction applied to the $V_\mathrm{r,HC}(\Phi_\mathrm{obs})$ is only meaningful for the objects that we end up classifying as CCs. 

The errors associated with the $V_{\gamma}$ determinations were calculated as the sum in quadrature of the error in the $V_\mathrm{r,HC}(\Phi_\mathrm{obs})$ from the cross-correlation and the error on the correction applied, $V_\mathrm{r,puls}(\Phi_\mathrm{obs})$. The error on this last term was estimated as the extreme values taken at the phase of the observation by the individual CCs used to build the given template. We added into this calculation a phase error to account for the uncertainty on the determination of the $\mathrm{HJD_0}$ for each star and the dispersion in the $\mathrm{Amp_{V_r}}$ derived from the relations presented in Fig.~\ref{fig:ampksrv}. This way we penalise with a higher error those stars that were observed in a phase where the pulsation velocity is changing faster or where the templates used for the correction are very different from each other.

We have two observations for each of B05 and D05, at different phases, that can give us a hint of the usefulness of our approach. We applied the RV correction independently to each of the observations. Before applying the pulsation correction, the individual RV measurements differ by 22.4\,\kms{} and 16.5\,\kms{}, respectively. As expected, after implementing this correction these differences diminish to 7\,\kms{} in both cases, and the values fall within the estimated errors.

\subsection{Cepheid kinematics and a comparison with the predicted LSR velocities}\label{subsection:Vlsr}

\begin{figure*}
\centering
    \includegraphics[width=17cm]{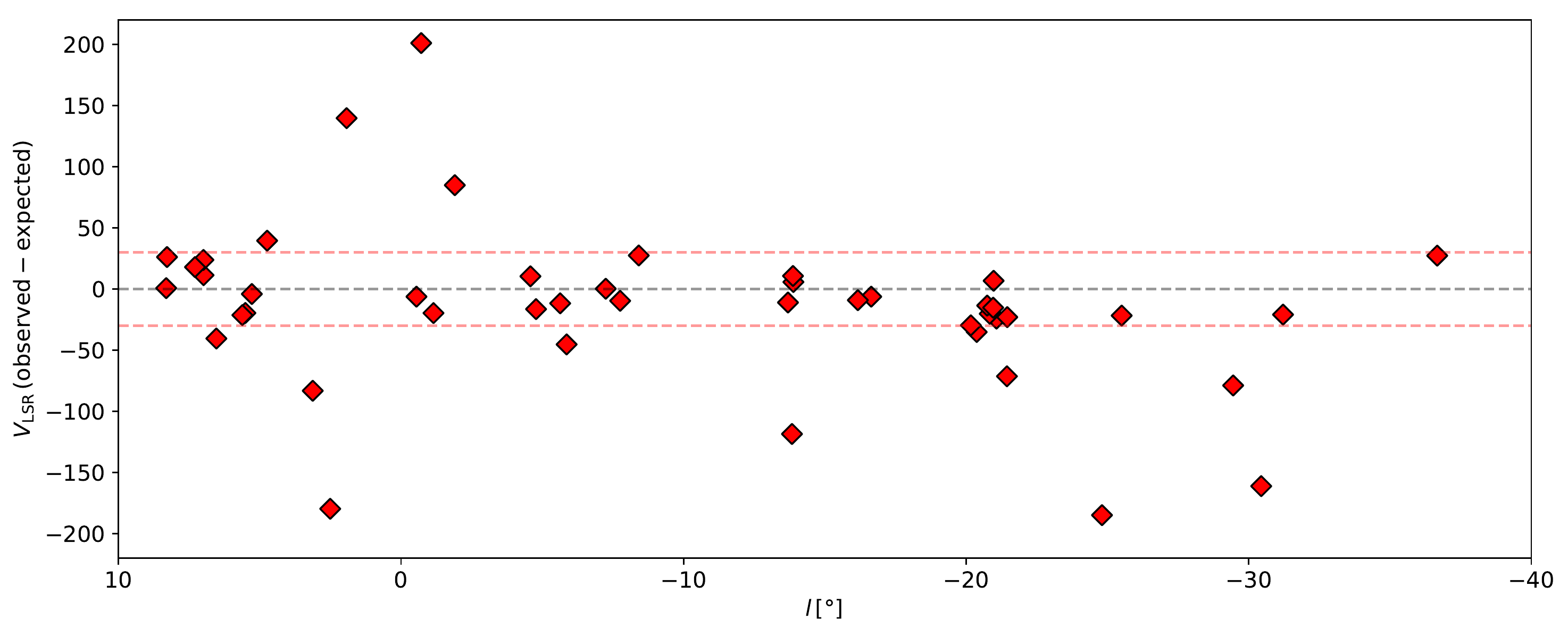}
        \caption{Difference between the observed $V_\mathrm{LSR}$ and the expected $V_\mathrm{LSR}$ as a function of Galactic longitude, for the Cepheid candidates in the bulge and disk regions. 
        The expected $V_\mathrm{LSR}$ values were estimated using the MW rotation curve determined by \citet{mroz2019}. For this purpose  the distances were calculated for each star using the  \citet{macri2015} PL relations (i.e., assuming that all of them are CCs; see text for details). Therefore, the observed velocities of those stars having large velocity differences would not be compatible with them actually being CCs. The horizontal red dashed lines indicate velocity differences of $\Delta V_\mathrm{LSR}=\pm30$\kms\ between the observed and expected values. The error bars for the measured velocities are, in general, lower than the point size in this figure.
        }

    \label{fig:rv_obs-exp}
\end{figure*}

In order to classify the Cepheid candidates, their observed local standard of rest (LSR) velocities, $V_\mathrm{LSR}$, were calculated and compared with the expected LSR velocities for disk stars at the same position
$(l, b, R_\mathrm{GC})$, where the Galactocentric distance, $R_\mathrm{GC}$, was estimated assuming they are CCs. In this way, we tested how compatible a star’s observed velocity was with the one it would have if it were a CC. The rotation curve from \citet{mroz2019} was adopted for this comparison. These authors use CCs to construct the rotation curve of the MW up to a distance from the Galactic center of $\sim20$\,kpc. This is the most accurate rotation curve available in the literature for Galactocentric distances higher than 12\,kpc and this work has shown the important role that CCs play in studying the rotation curve in the outer part of our Galaxy. The analytical form of the rotation curve is $\Theta(R) = \Theta_{0}+\frac{d\Theta}{dR}(R-R_{0}) = 233.6(\pm2.6)$\,\kms{}$-1.34(\pm0.20)$\,\kms \,kpc$^{-1}\,(R-R_{0})$, with $R_{0}$ the distance to the Galactic center. We adopted $R_{0}=8.122\pm0.031$\,kpc \citep{gravity2018}. We transformed the calculated heliocentric $V_{\gamma}$ of each Cepheid into the $V_\mathrm{LSR}$, assuming $(U_{\odot},V_{\odot},W_{\odot}) = (10.1,12.3,7.3)\,$\kms{}, which is the best-fit solar motion with respect to the LSR obtained by \citet{mroz2019} in their Model 2 rotation curve.

In Fig.~\ref{fig:rv_obs-exp} we show the comparison between the observed and expected $V_\mathrm{LSR}$ ($\Delta V_\mathrm{LSR}$), as a function of Galactic longitude, for the bulge and the disk samples. The final goal is to find CCs. From this plot we can already distinguish stars that show disk-like kinematics (and are therefore likely to be CCs) from those that have high $\Delta V_\mathrm{LSR}$ and are probably T2Cs that belong to the bulge, the halo, or the thick disk. We note that T2Cs are expected to have a large radial velocity dispersion, and therefore to show larger differences in $\Delta V_\mathrm{LSR}$, almost independently of their estimated distances, here calculated as if all of them were CCs. Nonetheless, we cannot exclude, based only on $\Delta V_\mathrm{LSR}$, that some T2Cs could accidentally have the same velocity of the disk in that position. This is the reason why we also use metallicities in what follows.

\section{Spectroscopic determination of atmospheric fundamental parameters}\label{section:classification}

In the previous section we obtain the systemic RVs for the CCs candidates. As discussed, this information is a useful complement that can already help us to find some contaminants, mainly T2Cs. Adding the atmospheric parameters for these stars is, as we show in this section, the best way to definitely isolate a clean sample of CCs.

A full spectral fitting technique was used to simultaneously derive fundamental stellar parameters of all the target stars: effective temperatures (\Teff{}), surface gravities (\logg{}), and metallicities (\feh). The analysis was restricted to the $H$- and $K$-band spectral ranges. In all cases, for the X-Shooter spectra we masked the wavelength regions shorter than $1.50 \, \mu {\rm m}$, $\sim 1.70 \rm{-} 1.97 \, \mu {\rm m}$, $\sim 2.00 \rm{-} 2.07 \, \mu {\rm m}$ and longer than $\sim 2.3 \, \mu {\rm m}$, mainly because they were heavily affected by telluric absorption.

\subsection{Spectral synthesis}

The grid of synthetic spectra was generated using the stellar spectral synthesis code {\tt SPECTRUM}\footnote{\url{http://www.appstate.edu/~grayro/spectrum/spectrum.html}}, written by Richard O. Gray \citep{gray1994} together with the ATLAS9 model atmospheres calculated by \citet{meszaros2012}, a compilation of the APOGEE $H$-band spectral line list constructed by \citet{shetrone2015}, and the $K$-band line list developed by \citet{thorsbro2018}. The spectral synthesis was done assuming solar chemical abundances from \citet{grevesse1998}.

The grid of synthetic spectra covers all the possible values of the atmospheric parameters found for Galactic CCs. In order to allow for T2Cs, we extend the grid down to $\mathrm{\feh} = -2.0\,\rm{dex}$. The microturbulent velocity was fixed to 4\,\kms{}, which is the mean value of the derived mean microturbulent velocity for a sample of well-known CCs studied by \citet{proxauf2018}.

\subsection{Full spectral fitting}\label{subsection:spectralfit}

The stellar atmospheric parameters of each star were determined by comparing the reduced, telluric corrected, continuum normalized near-IR spectra to the grid of synthetic spectra produced with {\tt SPECTRUM}. This step was done using the code {\tt FERRE} \citep{allendeprieto2006}, that interpolates within the grid and finds the best match to the observed spectra by minimizing the $\chi^2$.

Suitable spectral windows were manually selected for each target, where telluric absorption correction was of good quality and with a $\rm{S/N} \gtrsim$ 15-20. As an example, we show in Fig.~\ref{fig:spectrum} the best-fit synthetic spectrum obtained with {\tt FERRE} for one of our Cepheid candidates. The measured atmospheric parameters for both the bulge and disk samples are presented in Table~\ref{tab:bulgesampleparameters} and Table~\ref{tab:disksampleparameters}, respectively.

\begin{table}
\footnotesize
\caption{Atmospheric parameters for the XSL comparison stars.}
\centering
\begin{tabular}{lccccccc}
\hline \hline
ID          &\Teff          &   \logg       &     \feh          &   Ref.        \\
        &  K            &               &                   &                   \\
\hline
$\zeta$~Gem     &       5480        &   3.41        &   $-0.10$       &   this work   \\
        &   5180        &       1.4         &   $-0.19$     &         (1)   \\
        & $5602\pm32$   & $2.09\pm0.03$ &       $0.10$        &         (2)   \\
        &               &                   &$-0.11\pm0.10$ &         (3)   \\
        & $5494\pm7$ *  & $1.12\pm0.03$ &       $0.16\pm0.05$        &         (4)   \\
\hline
HD\,6229  &     5242        & 2.87          & $-1.08$         &   this work   \\
        &  $5200\pm150$ & $2.50\pm0.16$ & $-1.07\pm0.13$&         (5)   \\
        &  $5260\pm150$ & $2.55\pm0.3$  & $-1.02\pm0.15$&         (6)   \\ 
\hline
HD\,193896&     5300        &   3.22        &   $-0.27$       &   this work   \\
        &   4969        &   1.96        & $-0.17\pm0.12$&         (7)   \\ 
\hline
HD\,33299       &       4644        &   1.66        &   $-0.06$       &   this work   \\
        &   4626        &   1.50        &   0.26        &         (8)   \\
\hline
\end{tabular}
\tablebib{(1) \citet{romaniello2008}, (2) \citet{luck2011}, (3) \citet{genovali2014} (The iron abundance from \citet{romaniello2008} rescaled to their metallicity scale), (4) \citet{proxauf2018}, (5) \citet{for2010}, (6) \citet{afsar2012}, (7) \citet{luck2015}, (8) \citet{luck2014}. }
\tablefoot{{*}{Weighted mean of the \Teff, $\mathrm{\logg}$, and $\mathrm{\feh}$ values, in the range 5200-5850\,K, 0.4-1.9, and $-0.06$-0.39, respectively. }}
 \label{tab:XSL_stars}
\end{table}

\begin{table*}
 \footnotesize
 \caption{Radial
velocities and atmospheric parameters: Bulge sample.}
 \centering
 \begin{tabular}{lccccccl}
 \hline \hline
ID      &$V_\mathrm{r,HC}$&$V_\mathrm{\gamma}$  &       Period  &       \Teff   &     \logg       &  \feh         &       Type            \\
        &  \kms           &  \kms               &       days    &       K         &               &               &                       \\
\hline                                                                                                         
B01     &       $+51.2 $        &       $+43 \pm 4$     & 22.10 &       4969    &       2.5     &       $-0.11$         &               CC\tablefootmark{\;1}   \\
B02     &       $+145.9$        &       $+122\pm13$     & 25.72 &       5000    &       1.6     &       $-2.00$\tablefootmark{a} &       T2C\tablefootmark{\;2}  \\
B03     &       $-90.7 $        &       $-74\pm 11$     & 16.30 &       4396    &       2.0     &       $-0.65$         &               None\tablefootmark{*}\\
B04     &       $-28.9 $        &       $-10\pm  3$     & 11.58 &       5237    &       1.5     &       $-0.03$         &               CC\tablefootmark{\;1}   \\
B05     &       $+2.7  $        &       $-20\pm  5$     & 14.06 &       5338    &       2.5     &       $-0.27$         &               CC\tablefootmark{\;1}   \\
B05     &       $-19.7 $        &       $-13\pm  3$     & 14.06 &       5385    &       1.4     &       $-0.45$         &               CC\tablefootmark{\;1}   \\
B06     &       $+14.0 $        &       $-10\pm  2$     & 11.98 &       5061    &       2.2     &       $-0.19$         &               CC\tablefootmark{\;1}   \\
B07     &       $+15.2 $        &       $+22 \pm 3$     & 11.67 &       5043    &       1.8     &       $+0.15$         &               CC\tablefootmark{\;1}   \\
B08     &       $-29.6 $        &       $-18\pm  6$     & 10.15 &       5858    &       1.7     &       $+0.27$         &               CC\tablefootmark{\;2}   \\
B09     &       $-38.5 $        &       $-15\pm  4$     & 24.20 &       5330    &       0.7     &       $-0.38$         &               CC\tablefootmark{\;1}   \\
B10     &       $-62.0 $        &       $-52\pm  4$     & 37.35 &       4030    &       1.4     &       $-0.27$         &               None\tablefootmark{*}\\
B11     &       $+14.9 $        &       $-2 \pm  4$     & 11.23 &       5233    &       2.6     &       $+0.23$         &               CC\tablefootmark{\;1}   \\
B12     &       $+2.8  $        &       $+12 \pm 3$     & 11.22 &       5868    &       1.4     &       $+0.01$         &               CC\tablefootmark{\;1}   \\
B13     &       $-29.9 $        &       $-48\pm  3$     & 18.04 &       4270    &       2.2     &       $-0.33$         &               CC\tablefootmark{**}\\
B14     &       $-19.8 $        &       $+6 \pm  3$     & 15.95 &       5601    &       0.7     &       $-0.02$         &               CC\tablefootmark{\;1}   \\
B15     &       $-42.7 $        &       $-38\pm  3$     & 13.27 &       4681    &       0.7     &       $-0.09$         &               None\tablefootmark{*}\tablefootmark{1}\\
B16     &       $-0.4  $        &       $-7 \pm 14$     &  9.84 &       5915    &       0.7     &       $+0.22$         &               CC\tablefootmark{\;2}   \\
B17     &       $+20.7 $        &       $+2 \pm  5$     & 11.39 &       5153    &       2.0     &       $-0.18$         &               CC\tablefootmark{\;1}   \\
B18     &       $+36.5 $        &       $19 \pm  6$     & 13.59 &       4141    &       1.9     &       $-0.18$         &               None\tablefootmark{*}\\
B19     &       $-3.8  $        &       $-19\pm  2$     &  4.15 &       5610    &       2.5     &       $-0.02$         &               CC\tablefootmark{\;1}   \\
B20     &       $+78.9 $        &       $+74 \pm10$     & 15.58 &       5043    &       0.7     &       $-1.65$         &               T2C\tablefootmark{\;2}  \\
B20     &       $+73.9 $        &       $+85 \pm 7$     & 15.58 &       5043    &       1.0     &       $-1.58$         &               T2C\tablefootmark{\;2}  \\
B20     &       $+90.4 $        &       $+68 \pm 7$     & 15.58 &       5015    &       2.4     &       $-1.35$         &               T2C\tablefootmark{\;2}  \\
B21     &       $-0.6  $        &       $-27\pm  5$     & 13.52 &       5424    &       1.3     &       $-1.27$         &               T2C     \\
B22     &       $+203.6$        &       $+193\pm 7$     & 27.82 &       4956    &       0.6     &       $-1.59$         &               T2C         \\
B23     &       $-204.0$        &       $-199\pm 8$     & 34.36 &       3897    &       1.1     &       $-0.13$         &               None\tablefootmark{*}\\
B24     &       $-72.8 $        &       $-102\pm 7$     & 14.48 &       4779    &       1.2     &       $-1.68$         &               T2C\tablefootmark{\;2}  \\
\hline
 \end{tabular}
 \label{tab:bulgesampleparameters}
 
 \tablefoot{
 \tablefoottext{a}{Upper limit for \feh. The value for this object is at the lower \feh{} boundary of the grid of synthetic spectra.}
 \tablefoottext{1, 2}{Cepheid type as determined in \citet{dekany2019}, with 1 and 2 indicating that it was classified as a CC or T2C, respectively. }
 \tablefoottext{*}{\Teff{} too low to be a Cepheid (see text for explanation).}
 \tablefoottext{**}{\Teff{} at lower limit for a CC.}
 }
 \end{table*}

\begin{table*}
 \footnotesize
 \caption{Radial
velocities and atmospheric parameters: Disk sample.}
 \centering
 \begin{tabular}{lccccccl}
 \hline \hline
ID  &$V_\mathrm{r,HC}$&$V_\mathrm{\gamma}$&Period& \Teff & \logg &  \feh & Type   \\
    &  \kms           &  \kms             &days  &    K  &       &       &        \\
\hline                                                                                                              
D01 & $-9.3 $ & $+15   \pm 6$    & 17.48 & 4147 & 1.5  &  $-0.63$ &     None\tablefootmark{*}\tablefootmark{1}\\
D02 & $-97.1$ & $-111  \pm 8$    & 12.33 & 5158 & 1.7  &  $-0.85$ &     T2C\tablefootmark{\;2}          \\
D03 & $-20.0$ & $-6    \pm 8$    & 9.51  & 4879 & 3.0  &  $-0.55$ &     T2C      \\
D04 & $-15.3$ & $-35   \pm 8$    & 15.65 & 4964 & 2.1  &  $+0.13$ &     CC       \\
D05 & $-21.3$ & $-18   \pm 4$    & 9.98  & 5604 & 0.8  &  $+0.44$ &     CC       \\
D05 & $-4.8 $ & $-25   \pm 5$    & 9.98  & 5365 & 1.0  &  $+0.36$ &     CC       \\
D06 & $-15.3$ & $-24   \pm 2$    & 21.25 & 4850 & 1.5  &  $+0.11$ &     CC       \\
D07 & $+43.9$ & $+36   \pm 5$    & 18.53 & 4909 & 2.0  &  $-0.64$ &     CC       \\
D08 & $-13.3$ & $-24   \pm 5$    & 10.67 & 5328 & 2.3  &  $-0.11$ &     CC       \\
D09 & $+13.3$ & $+19   \pm 2$    & 8.45  & 5844 & 1.0  &  $-0.23$ &     CC\tablefootmark{\;1}    \\
D10 & $-53.6$ & $-69   \pm 4$    & 11.24 & 4439 & 1.8  &  $-0.51$ &     None\tablefootmark{*}\\
D11 & $+11.6$ & $+4    \pm 8$    & 8.13  & 5432 & 1.8  &  $-0.06$ &     CC       \\
D12 & $+20.2$ & $+8    \pm 5$    & 10.24 & 5166 & 3.2  &  $-0.11$ &     CC\tablefootmark{\;1}    \\
D13 & $-51.4$ & $-43   \pm 3$    & 9.73  & 5414 & 2.4  &  $-0.04$ &     CC       \\
D14 & $+20.1$ & $+12   \pm 6$    & 7.98  & 5598 & 2.4  &  $-0.17$ &     CC       \\
D15 & $-5.2 $ & $+0    \pm 4$    & 8.92  & 5384 & 1.9  &  $-0.15$ &     CC       \\
D16 & $-1.0 $ & $+13   \pm 9$    & 14.61 & 5896 & 0.5  &  $+0.23$ &     CC\tablefootmark{\;1}    \\
D17 & $+27.7$ & $+39   \pm 5$    & 10.24 & 5509 & 1.5  &  $-0.53$ &     CC\tablefootmark{\;2}    \\
D18 & $-73.8$ & $-90   \pm 8$    & 8.39  & 4715 & 2.3  &  $-0.70$ &     T2C      \\
D19 & $+11.5$ & $+25   \pm 5$    & 10.07 & 5240 & 2.0  &  $+0.03$ &     CC\tablefootmark{\;2}    \\
D20 & $+61.5$ & $+66   \pm 4$    & 7.74  & 5545 & 1.8  &  $-0.21$ &     CC\tablefootmark{\;1}    \\
D21 & $-84.8$ & $-77   \pm 3$    & 20.75 & 5370 & 0.3  &  $+0.23$ &     CC\tablefootmark{\;1}    \\
\hline                          
 \end{tabular}                  
 \label{tab:disksampleparameters}
 \tablefoot{
 \tablefoottext{1, 2}{Cepheid type as determined in \citet{dekany2019}, with 1 and 2 indicating that it was classified as a CC or T2C, respectively. }
 \tablefoottext{*}{\Teff{} too low to be a Cepheid (see text for explanation).}
 }
 \end{table*}

To the best of our knowledge there are not enough analyses available in the literature at the spectral resolution of this work and in the near-IR for the spectral types studied here. For this reason, we cannot provide a quantitative error estimate for the \feh{} values obtained for our objects. Instead, based on the variations seen on the fitted parameters while defining the spectral regions to be included in the final fit and our extensive experience in medium-resolution spectroscopy \citep[e.g., the GIBS survey:][]{vasquez2015,zoccali2017}, we conservatively estimate that the error on the measured $\mathrm{\feh}$ is $\approx \pm 0.2 \, \mathrm{dex}$, and therefore largely sufficient to discriminate between metal-poor T2Cs ($\mathrm{\feh} \lesssim -1$) and solar metallicity CCs.

In order to test the atmospheric parameter determination, we selected four stars from the X-Shooter Spectral Library \citep{chen2014}, a stellar spectral library comprised of X-Shooter spectra for more than 700 stars that cover a wide range of atmospheric parameters. The selected comparison objects were HD\,193896, $\zeta$~Gem (CC star), HD\,6229 (horizontal branch star), and HD\,33299. We retrieved from the archive the near-IR arm spectra for these objects that were taken with the same slit width, and thus have the same resolution as our targets. The selection of these benchmark stars was based on their parameters covering a range similar to that of our targets (they are GK giants) and also on the availability of accurate determinations of their atmospheric parameters in the literature. In Table \ref{tab:XSL_stars} we compare the atmospheric parameters derived in this work by means of the full spectral fitting technique in the $H$ and $K$ bands (as done for the sample of Cepheid candidates) with the most recent determinations available in the literature, which are based on optical, higher resolution spectroscopic analyses. We find that the \feh{} values are fully consistent within our posited precision of $0.2 \, \mathrm{dex}$.

We also verify for the stars that have more than one observation that the spread we see is $\sim0.2\,\rm{dex}$. In the case of B05 and D05, with both observations for each star having very similar S/N, the metallicities estimated are $\mathrm{\feh}\,(\Phi_{obs_1},\Phi_{obs_2})=(-0.27, -0.45) \, \mathrm{dex}$ and $\mathrm{\feh}\,(\Phi_{obs_1},\Phi_{obs_2})=(+0.36, +0.44) \, \mathrm{dex}$, respectively. For B20 we have three observations, with the first two having significantly lower S/N than the third. Even so, we find that the determined values, $\mathrm{\feh}\,(\Phi_{obs_1},\Phi_{obs_2},\Phi_{obs_3})=(-1.65, -1.58, -1.35) \, \mathrm{dex}$, fall within the quoted errors.

\subsection{Classification of Cepheid types}\label{subsection:classificationtype}

\begin{figure*}
\sidecaption
  \includegraphics[width=12cm]{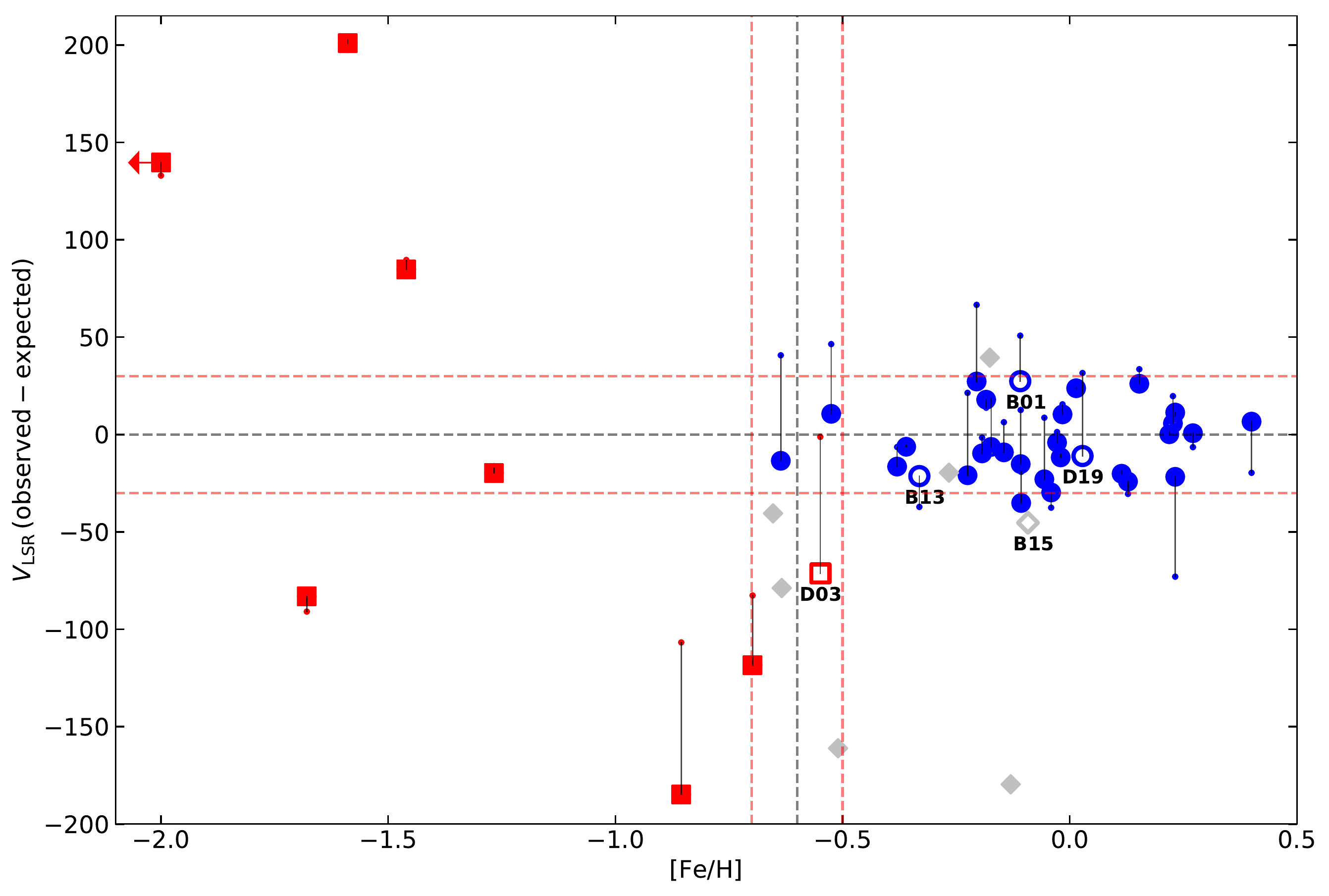}
     \caption{Difference between the observed $V_\mathrm{LSR}$ and the expected $V_\mathrm{LSR}$ ($\Delta V_\mathrm{LSR}$) as a function of the metallicity \feh, for the full sample of Cepheid candidates. The different markers indicate stars classified as CCs (blue circles), T2Cs (red squares), and low \Teff{} for a Cepheid (gray diamonds). See the text for details about the five stars represented with open symbols. The horizontal red dashed lines indicate velocity differences of $\Delta V_\mathrm{LSR}=\pm30$\kms\ and the black one $\Delta V_\mathrm{LSR}=0$\kms. The vertical black dashed line is $\mathrm{\feh}=-0.6\,\mathrm{dex}$ and the $\pm0.1\,\mathrm{dex}$ region is indicated with the vertical red dashed lines. There is a clear difference between CCs and T2Cs. Also included as a reference are the observed $V_\mathrm{LSR}$ on the y-axis (small points, connected with black lines to their $\Delta V_\mathrm{LSR}$ values). For the stars classified as CCs the dispersion around $\Delta V_\mathrm{LSR}=0$\kms\ decreases, thus compatible with having disk-like kinematics; instead, T2Cs show large $\Delta V_\mathrm{LSR}$.}
     \label{Vlsr-vs-FeH}
\end{figure*}

The measured \feh{} and $V_{\gamma}$ values presented in Table~\ref{tab:bulgesampleparameters} and Table~\ref{tab:disksampleparameters} allowed us to safely classify the Cepheid candidates. In Fig.~\ref{Vlsr-vs-FeH} we can clearly identify two groups of stars: a metal-rich sample that we identify as CCs, and a metal-poor sample that we classify as T2Cs. There is also a third group of stars that are too cold to be Cepheids. In what follows we discuss the groups separately.

\subsubsection{Classical Cepheids}\label{subsection:CCs}

As we have previously pointed out, CCs are young stars. Therefore, they are preferentially located in the thin disk, and are expected to be metal rich and to follow the rotation of the disk. In Fig.~\ref{Vlsr-vs-FeH} we plot the difference between the observed and expected velocity that each Cepheid candidate would have if it were a CC (as explained in Sect.~\ref{subsection:Vlsr}), as a function of \feh. From this figure, we can differentiate a metal-rich sample (30 objects with $\mathrm{\feh}\gtrsim -0.6\,{\rm dex}$) with $\Delta V_\mathrm{LSR}$ around zero, i.e., whose RVs behave as expected for stars in the disk following the MW rotation curve at their positions $(l,b,\mathrm{R_{GC}})$. We concluded that these are CCs (plotted as blue points in Fig.~\ref{Vlsr-vs-FeH}).

To clarify the impact in this plot of the assumption that all the stars are CCs, we also show their measured $V_\mathrm{LSR}$ before subtracting the expected value. As expected, CCs on the far side of the Galactic disk have $V_\mathrm{LSR}$ close to 0\,\kms\ since their motion is mainly perpendicular to our line of sight, as opposed to the T2Cs (red points in Fig.~\ref{Vlsr-vs-FeH}, discussed in Sect.~\ref{subsection:T2C}) that show a large velocity dispersion. It is also evident in this figure that, for the objects with $\mathrm{\feh}\gtrsim -0.6\,{\rm dex}$, the dispersion around $\Delta V_\mathrm{LSR}=0$\,\kms\ decreases when we subtract their corresponding expected velocities, meaning that they are compatible with their classification as CCs.

Within this group there are three stars (blue open circles in Fig.~\ref{Vlsr-vs-FeH}) for which we think it is worth doing a separate analysis. These are B01, B13, and D19. They have in common that they are among the farthest CCs in the sample, with $R_\mathrm{GC}$ between $\sim 23$ to 31\,kpc and at the same time most of them have solar \feh{} values, ranging from $+0.03$ to $-0.33\,\mathrm{dex}$. This can be unexpected at first glance since the MW disk is known to have decreasing metallicity with increasing Galactocentric distance. Even so, similar \feh{} values are found by \citet{genovali2014} at Galactocentric distances of 15-17\,kpc, where they show hints of the metallicity gradient having a large dispersion in the outer disk. Thus, B01 can be safely classified as a CC at $R_\mathrm{GC}\sim23$\,kpc, given that all its characteristics are fully consistent with this class, with $\mathrm{\feh}=-0.11 \, \rm{dex}$ and the difference between the observed and expected $V_\mathrm{LSR}$ being 28\,\kms{}. D19 is an interesting object; if it were a CC at $R_\mathrm{GC}\sim31$\,kpc, it would be tracing the young stellar populations of the MW beyond the limits usually set for the stellar disk. It is at the same time a difficult case since no comparison can be made with previously known behavior of CCs or thin disk rotation at such a long distance, simply because we do not have that information. Nonetheless, if we extrapolate the MW rotation curve up to its distance, we obtain an expected $V_\mathrm{LSR}=43$\,\kms{} that is compatible with the observed $V_\mathrm{LSR}=32 \pm 5$\,\kms{}. Considering that and its high metallicity, we classified this star as a CC. D19 is the farthest CC of the studied sample. A possible concern is that its extinction $A_{K_S}=1.40 \, \rm{mag}$ is not particularly high for a star at such a large distance. 

The star B13 has a low $\mathrm{\Teff{}}$ ($\sim4300\,\rm{K}$) value compared to what is usually found in spectroscopic studies of CCs analysing the behavior of $\mathrm{\Teff{}}$ as a function of phase. Recently \citet{proxauf2018} studied the phase variation in the atmospheric parameters of a sample of 14 well-known CCs, with periods ranging from $\sim3$ to 41\,days. None of the stars presented $\mathrm{\Teff{}}<4700\,\rm{K}$ at any of the observed phases. Phase-dependent studies were also done in a series of papers by \citet{luck2004}, \citet{kovtyukh2005}, \citet{andrievsky2005}, and \citet{luck2008} for CCs with periods from 3 to 68 days. No values of $\mathrm{\Teff{}}<4800\,\rm{K}$ were reported in any of these works. From a theoretical point of view, \citet{bono2000} have shown that their nonlinear pulsation models can get to lower temperatures than those discussed above, for periods $\gtrsim12$ days. In order to verify whether the low temperature measured for B13 ($\mathrm{P}= 18.04\,{\rm days}$) is consistent with its observed phase, we used nonlinear convective pulsation models \citep[see, e.g.,][]{marconi2005} computed for some of the period-$\mathrm{Amp}_{K_S}$ combinations of our observed Cepheid candidates to compare the observed and predicted \Teff{} at that particular phase. We found that B13 has in fact been observed at a minimum \Teff{} phase. Its temperature of $\sim4300\,\rm{K}$ is consistent with the predicted value of $\sim4400\,\rm{K}$ that we obtain from the closest pulsation model we have access to. We classified this star as a CC, although we warn the reader that such a low \Teff{} value is not commonly found in the literature. Observing this star at a different phase would be desirable to validate our classification.

Before we move onto the second group we note that there is a star, B15, that is plotted in Fig.~\ref{Vlsr-vs-FeH} (gray open diamond), which at first glance at its parameters  does not look different from the rest of the CCs described above. As we have just discussed, its derived $\rm{\Teff{}=4680\,K}$ is at the lower limit compared to temperatures measured for CCs in the literature. When looking at the possible temperature values for CCs with period $\sim13$ days, it is predicted that this low \Teff{} can be reached close to minimum \citep[see][]{bono2000}. However, we have found that B15 $\Phi_\mathrm{obs}$ corresponds to a maximum \Teff{} phase, thus inconsistent with its observed temperature. If it were a CC its distance from the Sun would be 31.5\,kpc, but it would have the same reddening as that obtained from the \citet{gonzalez2012} reddening map. As already mentioned, this map measures the reddening up to the distance of the bulge RC stars. Thus, it is not expected that this star at a distance $\sim4$ times greater would have a comparable reddening. Moreover, the difference between observed and expected $V_\mathrm{LSR}$ amounts to 45\,\kms{}. For it to be moving as expected from the rotation curve, this star should be at $d\lesssim14$\,kpc from the Sun. Given all the above considerations, we classify this star as belonging to the cold contaminants described in Sect.~\ref{subsection:cold}. 

\subsubsection{Type II Cepheids}\label{subsection:T2C}

A metal-poor ($ \mathrm{\feh}\lesssim-0.6\,\mathrm{dex}$) and high velocity dispersion group not following the disk rotation is also evident in Fig.~\ref{Vlsr-vs-FeH} (red squares). These are {bona fide} T2Cs. It is worth mentioning that most of them are in the Galactic bulge region. 

In the case of D03 (red open square in Fig.~\ref{Vlsr-vs-FeH}), we present a more detailed discussion of its classification because this star is at the \feh{} limit separating the two well-defined groups seen in Fig.~\ref{Vlsr-vs-FeH}. We note that if this star were a CC, its distance would be the greatest for the whole sample of CCs (45.2\,kpc from the Sun), while its extinction would be the lowest ($A_{K_S}=0.88 \, \rm{mag}$). In addition, as shown in this figure, its observed $V_\mathrm{LSR}$ is lower than expected for such a great distance, although this is not easy to assess since we do not have precise studies of the rotation curve at $R_\mathrm{GC}\sim40\,\rm{kpc}$ where this star would be located. Moreover, its distance assuming it is a T2C is 15.8\,kpc from the Sun and the expected $V_\mathrm{LSR}$ velocity at that distance ($+7$\,\kms) is in perfect agreement with the observed value ($-1\pm8$\,\kms). For all these reasons, we classify this star as a T2C that probably belongs to the thick disk on the far side of our Galaxy.

\subsubsection{Cold stars}\label{subsection:cold}

A third group also appears in Fig~\ref{Vlsr-vs-FeH} (gray diamonds). These six stars have $\Teff<4400\,\rm{K}$ and, together with B15, are too cold for any of the Cepheid classes. We have performed the same analysis as for B13 and B15, comparing their observed effective temperatures with the pulsation model values at $\Phi_\mathrm{obs}$, which are not compatible. These colder outliers may have been misclassified on the basis of their VVV light curves. The most plausible explanation is that these stars are ellipsoidal binaries, a close binary system whose brightness varies due to the presence of a red giant star deformed by tidal interaction with its companion \citep[see, e.g.,][]{pawlak2014}.

Given the short periods and large amplitudes of our cold objects, they should have high Roche lobe filling factors. Thus, their orbits would have been circularized by tidal forces and the resulting light curves would be sinusoidal and difficult to differentiate from Cepheids.

The contamination of CC candidate samples based on near-IR photometry by ellipsoidal variables has not been taken into account by previous works. This is probably because their amplitudes are typically lower than the amplitudes for CCs ($\mathrm{A}_I \lesssim 0.2 \, \mathrm{mag}$), but there are a few cases were they can be higher than $0.2 \, \mathrm{mag}$ and up to $0.3 \, \mathrm{mag}$ in the $I$ band \citep[e.g.,][]{pawlak2014}. We show here that they should be considered since they could be present in the same proportion as T2Cs\footnote{We consider it unlikely that these ``cool contaminants'' could be explained by a pointing mistake. The S/N in the spectra obtained for each of these sources is as expected given their magnitudes, and the probability of hitting by chance a star with such a similar brightness is very low, also considering that the pointing is done using a blind offset.}. 

\subsubsection{Comparison with previous works}

Recently, \citet{dekany2019} carried out a search of Cepheids using VVV near-IR photometry and classified them based on their light curve shapes into CCs and T2Cs using machine learning techniques. In this work, we have 26 stars in common that can be compared with their classification. These stars are flagged in column 8 of Tables~\ref{tab:bulgesampleparameters} and \ref{tab:disksampleparameters} with a superscript indicating the Cepheid class they were assigned to by \citet{dekany2019}. We note that there is, in general, good agreement with our work. From the 18 stars classified as CCs in their work, 16 are confirmed by us, while the other 2 (B15 and D01) were rejected due to their low measured \Teff{} values. The remaining eight variable stars in common were classified by \citet{dekany2019} as T2Cs. For this class, we find an agreement for only half of the objects, with four stars confirmed by us as T2Cs, while the remainder were re-classified as CCs based on their \feh{} and RV values. A larger sample of spectroscopically classified Cepheids is desirable to be able to further validate their photometric classification. It is also worth mentioning that there is one of our T2Cs (B02) that was previously classified in this class in the OGLE collection of variable stars \citep[see][]{soszynski2017}. Its ID in that catalogue is OGLE-BLG-T2CEP-662.

\subsection{Invisible cluster}

The Twin Cepheids presented in \citet{dekany2015a}, were also included in the present sample. These are two pulsating stars (IDs B11 and B12) with almost identical periods ($\sim11.2$ days), apparent magnitudes, and colors (thus identical extinctions) that are separated by just 18.3\arcsec, which corresponds to a projected distance of 1.46\,pc, considering the mean distance to the objects of $16.41\,\mathrm{kpc}$ measured here. These authors concluded that the pair of Cepheids were part of a star cluster invisible to us due to the high interstellar extinction. We confirm this scenario by comparing \feh{}\;values, $V_{\gamma}$, and distances (see Table~\ref{tab:twinceph}). According to our measurements, these two Cepheids have properties that allow us to confirm their classification as CCs and that are consistent with them belonging to the same stellar population.

\begin{table}
 \footnotesize
 \caption{Parameters of the Twin Cepheids.}
 \centering
 \begin{tabular}{lccccc}
 \hline \hline
ID      &$V_\mathrm{\gamma}$&   Period  &  \feh  & $\mathrm{d_{CC}}$ & $A_{K_S,\;\mathrm{CC}}$\\
    &  \kms                     &       days    &                & kpc               & mag\\
\hline                                                                                                         
B11     &       $-2 \pm  4$     & 11.23     &$0.23\pm0.20$&   $16.63 \pm 0.74$ & 2.48    \\
B12     &       $+12 \pm 3$     & 11.22     &$0.01\pm0.20$&   $16.18 \pm 0.75$ & 2.51    \\
\hline
 \end{tabular}
 \label{tab:twinceph}
 \end{table}

\subsection{Metallicity gradient}\label{subsection:FeH_gradient}

\begin{figure*}
\sidecaption
  \includegraphics[width=12cm]{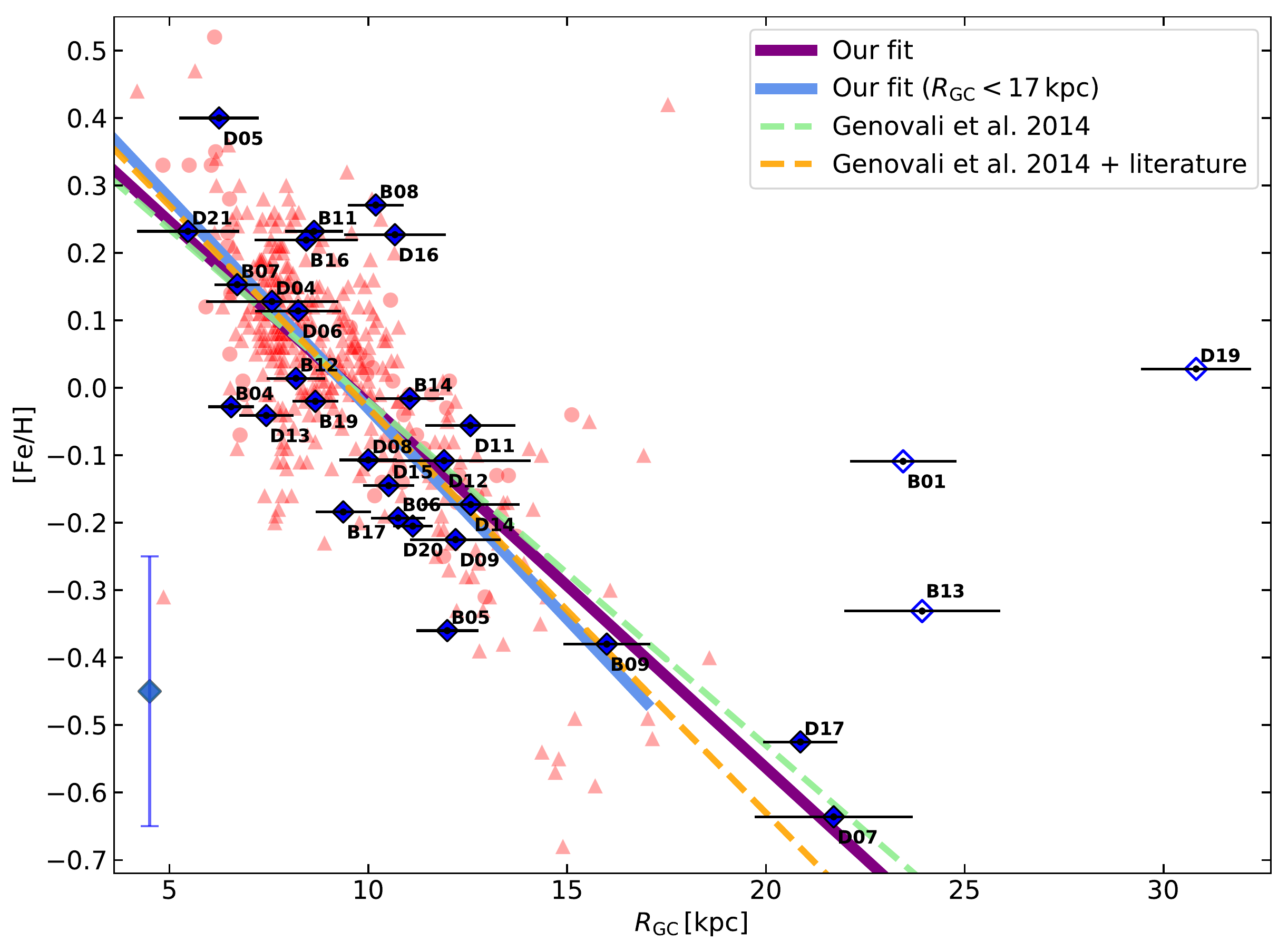}
     \caption{Iron abundance for the far disk CCs classified in this work as a function of their Galactocentric distance, plotted using filled blue diamonds. Open symbols are used for the same stars as in Fig.~\ref{Vlsr-vs-FeH}. The distances have been computed assuming $A_{K_S}/E(J-K_S)=0.465$ and $A_{K_S}/E(H-K_S)=1.308$. The error bars in the distances represent the total uncertainties calculated taking into account all the sources of uncertainty discussed in Sect.~\ref{subsection:reddening_law}. The metallicity error bar is shown in the bottom left corner of the plot ($\sigma_{\feh{}}=0.2 \, \rm{dex}$, see Sect.~\ref{subsection:spectralfit}). The metallicity gradient measured by \citet{genovali2014}, based only on abundances provided by their group (red light circles), is shown as a light green dashed line; the orange dashed line is their fit including iron abundances for CCs available in the literature, which are plotted with red light triangles. The purple and light blue lines show the metallicity gradient on the far side of the Galactic disk as derived in this work, when using CCs in the whole range of $R_\mathrm{GC}$ and for $R_\mathrm{GC}\lessapprox\,17\,\rm{kpc}$, respectively. }
     \label{fig:FeH_gradient_new}
\end{figure*}

Combining the measured distances and metallicities presented in Tables~\ref{tab:bulgesamplespectralinfo},~\ref{tab:disksamplespectralinfo},~\ref{tab:bulgesampleparameters}, and~\ref{tab:disksampleparameters} we can also trace the radial metallicity gradient of the young stellar population in the other side of the MW disk. This is shown in Fig.~\ref{fig:FeH_gradient_new}, where we plotted the bulge and disk samples, together with the gradients derived in the near side of the disk by \citet{genovali2014} with and without literature data. We fitted the metallicity gradient, considering the uncertainties in both $R_\mathrm{GC}$ and \feh{}, and determined a slope of $-0.062\pm0.013\,\mathrm{dex\, kpc^{-1}}$ with an intercept of $+0.59\pm0.13\,\rm{dex}$ when considering CCs with $R_\mathrm{GC}\lessapprox\,17\,\rm{kpc}$ (the values of the slope and intercept are $-0.054\pm0.008\,\mathrm{dex\, kpc^{-1}}$ and $+0.52\pm0.08\,\rm{dex}$, respectively, when the fit is done for the whole sample of CCs). We note that our sample extends the study of the metallicity distribution up to $R_\mathrm{GC}\sim24\,\mathrm{kpc}$ and is in very good agreement with the relations found by \citet{genovali2014}.

It can be seen in Fig.~\ref{fig:FeH_gradient_new} that for $R_\mathrm{GC} \gtrsim 20$\,kpc there are some solar metallicity ($\mathrm{\feh}\sim-0.1$) stars that do not follow the main trend seen for the full sample. They could be a hint of the existence of a flattening in the metallicity gradient. A similar behavior is observed in \citet{genovali2014} (their Fig. 4) at $R_\mathrm{GC}$ between $\sim$14 and $\sim$17 \,kpc for a few stars, although in both cases the number of CCs at such large distances is scarce. The authenticity of such flattening is worth further investigation.

We note that there is a gap in our data, with only one star in the range of Galactocentric distances 13-20\,kpc. We think this is a selection effect caused by a scarcity of CCs with periods longer than about 5 to 6 days, i.e., the youngest CCs, at great Galactocentric distances, as shown by \citet{skowron2019} in their Figs. 3 and S4. The periods of the Cepheids in our sample are $\gtrapprox$ 8 days with only one exception with $P\approx4.15$ days (see Tables \ref{tab:bulgesample} and \ref{tab:disksample}).

   \begin{figure}
   \centering
   \resizebox{\hsize}{!}{\includegraphics{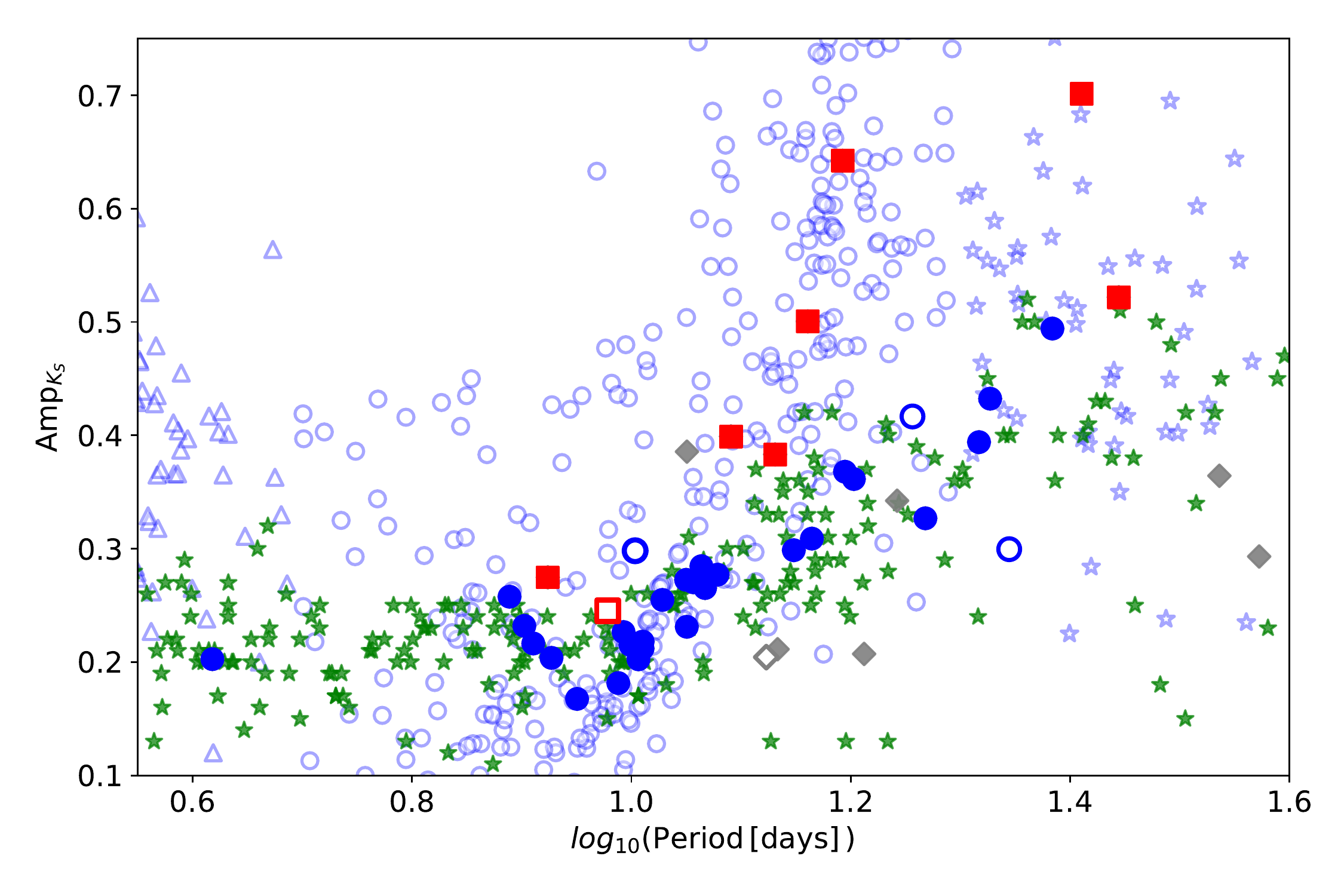}}
      \caption{Position of the sample of candidate Cepheids in the  Bailey diagram, now classified as CCs (blue circles) and T2Cs (red squares). In the background, other well-known CCs (light green stars) and T2Cs in the MW (light blue  open symbols) are plotted. Literature T2Cs are divided into the three main subtypes  described in Fig.~\ref{Bailey}. The gray diamonds indicate stars whose \Teff{} values are too low to be Cepheids. 
              }
         \label{Bailey_classified}
   \end{figure}

\section{Summary and conclusions}\label{section:summary}

We obtained the largest sample of spectroscopically confirmed CCs on the far side of the MW disk. We classified the sample stars into Classical and type II Cepheids using single-epoch near-IR X-Shooter spectra to distinguish between metal-rich CCs with disk-like kinematics and metal-poor T2Cs characterized by higher velocity dispersion. Out of a total of 45 stars, we found 30 CCs and 8 T2Cs. The remaining seven objects were probably misclassified as Cepheid candidates because their spectroscopic temperatures are too low to be Cepheids, and are likely to be ellipsoidal binaries instead. By means of the present data, we have increased the sample of spectroscopically confirmed CCs on the far side of the disk by a factor of $\sim10$.

We demonstrated that near-IR ($H$- and $K$-band) spectroscopy at $R \sim 8000$ (X-Shooter@VLT) allows us to unambiguously distinguish between Classical and type II Cepheids based on their radial velocity and metallicity (and \Teff{}) information, when coupled with their near-IR photometry coming from the VVV survey. As expected \citep[see, e.g.,][]{soszynski2017}, most T2Cs are located in the region of the bulge. The location of both types of Cepheids in the Bailey diagram (see Fig.~\ref{Bailey_classified}) is also fully consistent with the distribution for well-known Classical and type II Cepheids.

We determine for the first time the metallicity gradient in a region of our Galaxy that has remained largely unexplored. It is fully consistent with the gradient traced by \citet{genovali2014} using CCs on the near side of the disk. This is quite reassuring, considering the lower resolution of the spectra used in our study, and the larger distance uncertainty affecting our CCs. The main source of uncertainty in the distances, when studying objects at low Galactic latitudes and towards the inner regions of the Galaxy, is the lack of a well-constrained infrared reddening law available in the literature. Small changes on its value significantly impact the measured distances of these highly reddened objects. Because we have near-IR photometry for a sample of CCs, and we know their intrinsic colors very accurately from their PL relations, we used them to provide a measurement of the total-to-selective extinction ratios, $A_{K_S}/E(J-K_S)=0.465$ and $A_{K_S}/E(H-K_S)=1.308$, which are both in good agreement with previous studies.

Given the results of the present analysis, both the identification of new Cepheids close to the Galactic plane and their spectroscopic follow-up should be encouraged. The far side of the Galaxy is not out of reach anymore.

\begin{acknowledgements}
We gratefully acknowledge data from the ESO Public Survey program ID 179.B-2002 taken with the VISTA telescope, and products from the Cambridge Astronomical Survey Unit (CASU) and from the VISTA Science Archive (VSA). Support for this work was provided by the BASAL Center for Astrophysics and Associated Technologies (CATA) through grant AFB170002, and the Ministry for the Economy, Development and Tourism, Programa Iniciativa Cient{\'i}fica Milenio grant IC120009, awarded to the Millennium Institute of Astrophysics (MAS). This work is part of the Ph.D. thesis of J.H.M., funded by grant CONICYT-PCHA Doctorado Nacional 2015-21151640. We acknowledge additional support by Proyecto Fondecyt Regular \#1191505 (M.Z.), \#1171273 (M.C.), \#1170121 (D.M.). A.R.A acknowledges partial support from FONDECYT through grant \#3180203. We thank I. Dékány and G. Hajdu for providing the corrected aperture photometry for the disk sample. This research has made use of NASA’s Astrophysics Data System Bibliographic Services, the SIMBAD database and the ``Aladin sky atlas'' developed at CDS, Strasbourg Observatory, France. 
This work made use of PYTHON routines in the {\tt astropy} package \citep{astropy13}.
\end{acknowledgements}


\bibliographystyle{aa}
\bibliography{biblio}

\end{document}